\documentclass[conference]{IEEEtran}

\IEEEoverridecommandlockouts

\usepackage[hidelinks]{hyperref}

\usepackage{wrapfig}
\usepackage{verbatim} 
\usepackage{amsmath,amssymb,amsthm} 
\usepackage{listings} 
\usepackage{color}
\usepackage[dvipsnames]{xcolor}

\usepackage{graphicx}

\usepackage{enumerate}
\usepackage{paralist}

\usepackage[oldcommands,linesnumbered,boxed]{algorithm2e}

\setalcapskip{1em} 
\newcommand{\martinAlgFontsize}{\fontsize{8.0}{8.0}\selectfont}
\SetAlFnt{\martinAlgFontsize}
\newcommand{\martinAlgCapFontsize}{\fontsize{9.0}{9.0}\selectfont}
\SetAlCapNameFnt{\martinAlgCapFontsize}
\SetAlCapFnt{\martinAlgCapFontsize}

\newcommand{\martinTableFontsize}{\scriptsize\selectfont}

\newcommand{\martinListingCodeFontsize}{\fontsize{5.5}{7.5}\selectfont}
\newcommand{\martinListingFontsize}{\fontsize{8}{8}\selectfont}

\SetKwInOut{Global}{Global}
\SetKwInOut{Name}{Name}

\newcommand{\ackContent}{
    This work has been supported by the European Union's Horizon 2020 research and innovation programme under grant No 739551 (KIOS CoE).
}

\newcommand{\andnodes}{\Delta} \newcommand{\ornodes}{\Theta}
\newcommand{\degin}{d_{in}} \newcommand{\degout}{d_{out}}

\newcommand{\gdomain}{\mathcal{G}}
 
\newcommand{\cost}{\varphi} 
\newcommand{\costprime}{\varphi'} 
\newcommand{\form}{f_{G}} 

\newcommand{\concat}{\cdot} 

\newcommand{\wcc}{wcc} 
\newcommand{\rem}{\sigma} 
\newcommand{\tool}{META4ICS } 

\newcommand{\budget}{\gamma} 

\colorlet{punct}{red!60!black}
\definecolor{background}{HTML}{FFFFFF}
\definecolor{delim}{RGB}{20,105,176}
\colorlet{numb}{magenta!60!black}

\lstdefinelanguage{json}{
	basicstyle=\normalfont\ttfamily,
	numberstyle=\scriptsize,
	stepnumber=1,
	numbersep=8pt,
	showstringspaces=false,
	breaklines=true,
	frame=lines,
	stringstyle=\color{blue},
	backgroundcolor=\color{background},
	literate=
	*{0}{{{\color{numb}0}}}{1}
	{1}{{{\color{numb}1}}}{1}
	{2}{{{\color{numb}2}}}{1}
	{3}{{{\color{numb}3}}}{1}
	{4}{{{\color{numb}4}}}{1}
	{5}{{{\color{numb}5}}}{1}
	{6}{{{\color{numb}6}}}{1}
	{7}{{{\color{numb}7}}}{1}
	{8}{{{\color{numb}8}}}{1}
	{9}{{{\color{numb}9}}}{1}
	{:}{{{\color{punct}{:}}}}{1}
	{,}{{{\color{punct}{,}}}}{1}
	{\{}{{{\color{delim}{\{}}}}{1}
	{\}}{{{\color{delim}{\}}}}}{1}
	{[}{{{\color{delim}{[}}}}{1}
	{]}{{{\color{delim}{]}}}}{1},
}

\def\BibTeX{{\rm B\kern-.05em{\sc i\kern-.025em b}\kern-.08em
    T\kern-.1667em\lower.7ex\hbox{E}\kern-.125emX}}
\begin{document}

\thispagestyle{plain}
\pagestyle{plain}

\title{Identifying Security-Critical Cyber-Physical Components in Industrial Control Systems}

\author{   
	\thanks{\ackContent}
	\IEEEauthorblockN{
		Mart\'in Barr\`ere\IEEEauthorrefmark{1},         
		Chris Hankin\IEEEauthorrefmark{1}, 
		Nicolas Nicolaou\IEEEauthorrefmark{2}, 
		Demetrios G. Eliades\IEEEauthorrefmark{2}, 
		Thomas Parisini\IEEEauthorrefmark{3}		
	}
	\IEEEauthorblockA{\IEEEauthorrefmark{1}Institute for Security Science and Technology, Imperial College London, UK
		\\\{m.barrere, c.hankin\}@imperial.ac.uk}
	\IEEEauthorblockA{\IEEEauthorrefmark{2}KIOS Research and Innovation Centre of Excellence, University of Cyprus
		\\\{nicolasn, eldemet\}@ucy.ac.cy}		
	\IEEEauthorblockA{\IEEEauthorrefmark{3}Department of Electrical and Electronic Engineering, Imperial College London, UK
		\\\{t.parisini\}@imperial.ac.uk}		
}

\maketitle

\begin{abstract}
In recent years, Industrial Control Systems (ICS) have become an appealing target for cyber attacks, having massive destructive consequences. Security metrics are therefore essential to assess their security posture. In this paper, we present a novel ICS security metric based on AND/OR graphs that represent cyber-physical dependencies among network components. Our metric is able to efficiently identify sets of critical cyber-physical components, with minimal cost for an attacker, such that if compromised, the system would enter into a non-operational state. We address this problem by efficiently transforming the input AND/OR graph-based model into a weighted logical formula that is then used to build and solve a Weighted Partial MAX-SAT problem. Our tool, META4ICS, leverages state-of-the-art techniques from the field of logical satisfiability optimisation in order to achieve efficient computation times. Our experimental results indicate that the proposed security metric can efficiently scale to networks with thousands of nodes and be computed in seconds. In addition, we present a case study where we have used our system to analyse the security posture of a realistic water transport network. We discuss our findings on the plant as well as further security applications of our metric. 
\end{abstract}


\newcommand{\content}{sections} 

\section{Introduction}
\label{sec:intro}

From water and energy plants, to oil, gas, power, manufacturing, and automotive facilities, Industrial Control Systems (ICS) have become an appealing target for attackers over the last \hbox{years \cite{Humayed2017}}. 
Reasons for that include mostly their increased connectivity to the outside world, their lack of preparedness for cyber attacks, and the huge impact these attacks may have on many aspects of modern society. As a vital part of critical national infrastructure, protecting ICS from cyber threats has become a high priority since their compromise can result in a myriad of different problems, from service disruptions and economical loss, to jeopardising natural ecosystems and putting human lives at risk.  
Stuxnet, Industroyer, NotPetya, and more recently, WannaCry, exemplify the devastating consequences this type of attack may have on critical ICS infrastructures \cite{CyberXReport2019}, \cite{Lee2016}, \cite{Falliere2011}. 
In particular, cyber attacks on these systems can lead, for example, to flooding, blackouts, or even nuclear disasters \cite{Humayed2017}.

\begin{figure}[!t]
    \centering
    \includegraphics[scale=0.70]{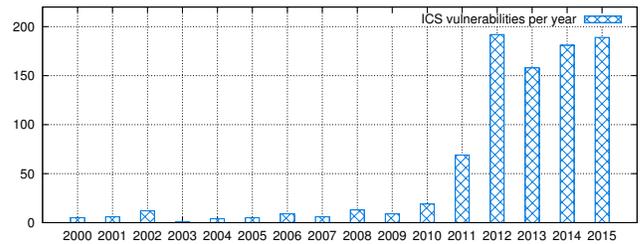}
    \caption{ICS vulnerabilities by year \cite{Kaspersky2016}}
    \label{fig:vulns_stats}
\end{figure}

In recent years, the number of vulnerabilities affecting ICS environments has dramatically increased. Figure \ref{fig:vulns_stats} shows this behaviour between 2000 and 2015 \cite{Kaspersky2016}. In 2015, almost a half of the ICS security flaws were classified as high risk vulnerabilities based on their CVSS v3 scores \cite{CVSS}. 
In 2017, a survey conducted over 21 countries, including a considerable number of ICS companies, shows that 54\% of them experienced at least one ICS security incident over the previous 12 months \cite{BusinessAdvantage2017}. For large companies (500+ employees), the annual cumulative loss is reported to be \$497,097. During 2017-2018, an independent study conducted over 850 production ICS networks indicates that, among many alarming findings, 69\% of the sites have plain-text passwords traversing their ICS networks, and 53\% of them still use obsolete Windows systems such as Windows XP \cite{CyberXReport2019}. Although guidance and standard best practices are available to increase ICS security \cite{Stouffer2015}, the amount of cyber incidents just keeps increasing \cite{Threatscape2018}. 

\begin{figure}[!t]
	\centering
	\includegraphics[scale=0.35]{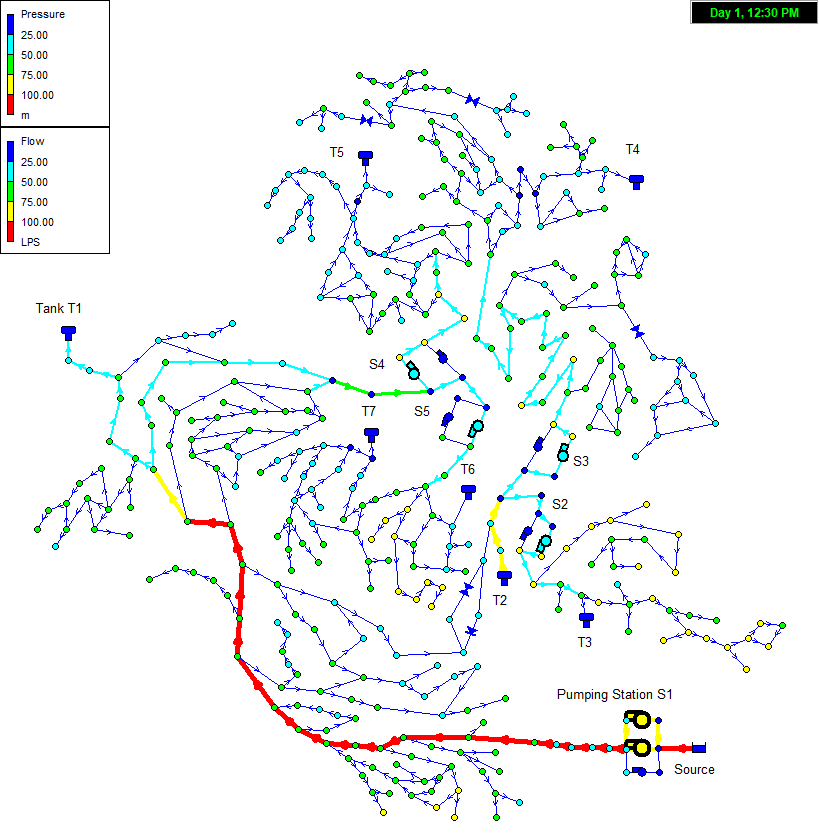}
	\caption{Large-scale Water Distribution Network with 7 tanks, 11 pumps and 4 valves. The system can be operated with 13 PLCs (C-Town Benchmark)~\cite{Ostfeld2012}}
	\label{fig:large-scale-wdn}
\end{figure}

These numbers come as no surprise since ICS environments, originally designed to work in isolation, suddenly became immersed into a hyper-connected world, just a few commands away from malicious actors. We argue that the integration of these complex environments, involving tangled ensembles of dependencies between cyber-physical components, has produced convoluted ecosystems that are hard to control and protect. As an example, Figure \ref{fig:large-scale-wdn} shows an open benchmark water distribution network that resembles a real city \textit{C-Town}~\cite{Ostfeld2012}, and illustrates the scale and structural complexity of these networks (discussed later in the paper). In that context, security metrics play a fundamental role since they allow us to understand the exposure and vulnerability of ICS environments, and improve their security posture \cite{Stouffer2015}. In particular, the ability to measure the security level of ICS environments and identify critical cyber-physical components that should be prioritised and addressed from a security standpoint becomes essential.  

In this paper, we propose an efficient and scalable security metric intended to identify critical cyber-physical components in ICS networks. Our work extends the methodology presented in \cite{Nicolau2018} where the security level of a system is understood as the minimum number of CPS components an attacker must compromise in order to disrupt the operation of the system. In addition, when each CPS component involves a \textit{cost of compromise}, our security metric is also able to identify the set of critical components with minimal cost for the attacker such that their compromise can take the system to a non-operational state. Our metric relies on AND/OR graphs that represent cyber-physical dependencies among network components and leverages state-of-the-art techniques from the field of logical satisfiability optimisation to achieve efficient computation times. Our experimental results indicate that the proposed security metric can efficiently scale to networks with thousands of nodes and be computed in seconds.  

\subsection{Method summary} 
Roughly stated, our approach takes as input an AND/OR graph, in the form of a digraph $G=(V,E)$ that represents the operational dependencies of the ICS environment, and a target node $t$. 
Then, we transform the graph into an equivalent logical formula that fulfils node $t$. 
The negation of this formula corresponds to the objective of the attacker, i.e. to disable $t$, which is later converted to an equisatisfiable formula in conjunctive normal form (CNF) by using the Tseitin transformation \cite{Tseitin70}. 
Finally, the CNF formula is used as a basis to build a Weighted Partial MAX-SAT problem. The solution to this problem is a minimal weighted vertex cut in $G$ with regards to node $t$. 
This cut represents the set of critical components with minimal cost for the attacker such that, if compromised, it will render the cyber-physical system into a non-operational state.

\subsection{Contributions}
Our main contributions are: 
\begin{inparaenum}[(1)]
    \item a mathematical model able to represent complex dependencies in ICS environments,
    \item a novel security metric and efficient algorithms to identify critical cyber-physical components, 
    \item a tool to analyse real ICS models and an extensive set of experiments on performance and scalability, and 
    \item a comprehensive case study conducted on a subset of a real water transport network that shows the applicability and further uses of our security metric. 
\end{inparaenum}

\subsection{Organisation of the paper}
Section \ref{sec:rw} discusses related work and existing limitations to identify critical components in ICS environments.  
Section \ref{sec:model} describes our graph-based modelling of ICS networks whereas 
Section \ref{sec:metric} formalises the proposed security metric. 
Section \ref{sec:computation-strategy} describes the steps to compute our security metric whereas  
Section \ref{sec:impl} describes our tool (META4ICS) and exemplifies the overall process. 
Section \ref{sec:analytical-exp} shows our experimental results in terms of performance and scalability whereas 
Section \ref{sec:case_study} details a case study and further uses of our security metric. 
Section \ref{sec:conclusions-fw} presents conclusions and future work.

\section{Related work}
\label{sec:rw}

Since the early 2000s, many research efforts have been produced to understand and improve the security of industrial control systems and national critical infrastructure \cite{Desmedt2002}, \cite{Desmedt2004}. 
These works have inspired the need for taking into account the cyber-physical dependencies between ICS components and being able to combine them in order to provide quantifiable measurements \cite{Humayed2017}, \cite{Hankin2016}. 
As such, the present work considers the insightful methodology presented in \cite{Nicolau2018}. 
However, the latter does not consider concrete algorithms to identify critical CPS components in ICS networks nor a model able to capture logical combinations among cyber-physical components with AND/OR connectives. 
This paper extends these ideas by proposing a complete AND/OR graph-based modelling capable of grasping complex interdependencies among CPS components. 
 
From a graph-theoretical perspective, our security metric looks for a minimal weighted vertex cut in AND/OR graphs.   
This is an NP-complete problem as shown in \cite{Desmedt2002, Desmedt2004, DesmedtUsingFF, Souza2013}. 
While well-known algorithms such as Max-flow Min-cut \cite{Ford1962, Dantzig1955} and variants of it could be used to estimate such metric over OR graphs in polynomial time, their use for general AND/OR graphs is not evident nor trivial as they may fail to capture the underlying logical semantics of the graph. 
In that context, we take advantage of state-of-the-art techniques which excel in the domain of logical satisfiability and boolean optimisation problems \cite{Davies2011}. 

A close research area to our problem includes the domain of attack graphs \cite{Barrere:CNSM2017}, \cite{Barrere:Naggen2017}, \cite{WangAlbaneseJajodia2014}, \cite{Shandilya2014}, \cite{Bopche2014}, \cite{Singhal2011NIST}. 
While attack graphs are mainly focused on depicting the many ways in which an attacker may compromise assets in a computer network, our approach is essentially different as we consider that network nodes can be equally compromised. In addition, attack graphs usually take into account only cyber lateral movements and do not consider cyber-physical dependencies among components~\cite{Humayed2017}. Moreover, real ICS models based on AND/OR graphs might also be cyclic, thus presenting the interdiction problem \cite{Altner2010}. We take a similar approach to that considered in \cite{Wang2017} in order to deal with cycles. 

Other attempts to identify critical cyber-physical components have been made in the domain of network centrality measurements \cite{Deng2018}. 
While useful in many type of scenarios \cite{Steiner2018}, almost all of them are focused on OR-only graph-based models for IT networks. 
In addition, we realise that automating ICS asset mapping is not an easy task for different reasons, among these, because active probing and scanning may be too intrusive, which might raise concerns about operational disruptions.  
However, understanding the full cyber-physical ecosystem is vital to maintain healthy Operational Technology~(OT) networks \cite{Ginter2019}, \cite{CyberXReport2019}. 
This is a premise that many security platforms already take into account, e.g., in the form of passive monitoring \cite{Nessus, CyberXAssessmentTool2018}. 
In the next section, we introduce our graph-based modelling approach for ICS environments.  
\section{ICS network modelling}
\label{sec:model}

We model an industrial network $W$ as a directed graph $G = (V,E)$ that represents the operational dependencies in $W$. 
\subsection{Graph modelling}
The graph involves three types of basic vertices, called atomic nodes ($V_{AT}$), that model the different classes of components in the network: 
	
	\begin{itemize}
		\item $S$ represents the set of sensor nodes in the network, 
		\item $C$ represents the set of actuator nodes, 
		\item $A$ represents the set of software agents. 
	\end{itemize}

We define $V_{AT} =  S \cup C \cup A$. In addition, the graph also involves two artificial node types that are used to model logical dependencies between network components: 
	\begin{itemize}
	\item $\andnodes$ represents the set of logical AND nodes, and   
	\item $\ornodes$ represents the set of logical OR nodes. 
\end{itemize}

Based on the previous types, we have that $V(G) = V_{AT} \cup \andnodes \cup  \ornodes$. 

$E(G)$ corresponds to the set of edges among nodes and their semantics depend on the type of nodes they connect. 
We consider three types of basic edges as follows: 
	\begin{itemize}
	\item $E_{A,C}=\{(a,c) : a \in A \land c \in C\}$ represents that agent $a$ controls the operation of actuator $c$, 
	\item $E_{S,A}=\{(s,a) : s \in S \land a \in A\}$ means that agent $a$ requires measurements from sensor $s$ to fulfil its purpose, 
	\item $E_{A,A}=\{(a_i,a_j) : a_i, a_j \in A, a_i \neq a_j\}$ represents that agent $a_j$ requires input from agent $a_i$ to operate normally. 
\end{itemize}
 
In addition, we consider another two types of edges involving artificial logical nodes. 
In particular, AND and OR nodes act as special connectors and shall be interpreted from a logical perspective. A node $v$ reached by an OR node means that the operational purpose of $v$ can be satisfied, i.e. $v$ operates normally, if \textit{at least one} of the incoming nodes to the OR node is also satisfied. Alike, a node $w$ reached by an AND node will be satisfied if \textit{all} of the incoming nodes to the AND node are also satisfied. 
The following two edge types are also allowed in $E(G)$: 
\begin{itemize}	
	\item $E_{i\andnodes\ornodes}=\{(v,x) : v \in V-C, x \in \andnodes \cup \ornodes\}$ represents incoming connections to AND/OR nodes from any type of graph node except actuators ($C$), 
	\item $E_{o\andnodes\ornodes}=\{(x,v) : x \in \andnodes \cup \ornodes, v \in V-S\}$ represents outgoing connections from AND/OR nodes to any type of graph node except sensors ($S$).
\end{itemize}

\subsection{Graph properties}
We use $\gdomain$ to denote the domain of AND/OR graphs. 
Let $\degin:V \rightarrow \mathbb{N}$ and $\degout:V \rightarrow \mathbb{N}$ be two functions that compute the in-degree and out-degree of a node respectively. 
The following properties hold in every instance of $G=(V,E)$, with $G \in \gdomain$: 

\begin{itemize}
	\item $\degout(c) = 0, \forall c \in C$ (no outgoing edges from actuators)
	\item $\degin(v) = 1, \forall v \in V_{AT}$ (only one incoming edge on atomic nodes)
	\item $\degout(v) \geq 1, \forall v \in \ornodes \cup \andnodes$ (logical nodes must connect to some destination node)
	\item $\degin(v) \geq 2, \forall v \in \ornodes \cup \andnodes$ (logical nodes combine two or more nodes)
\end{itemize}

\vspace{-0.2cm}
\subsection{Adversarial model}
Our adversarial model considers that an attacker can compromise any network node $n \in V_{AT}$ at a certain cost $\cost(n)$, with $\cost : V_{AT} \rightarrow \mathbb{R}_{\geq0}$. A \textit{compromised node} in this context shall be understood as a CPS component unable to operate properly, that is, a node incapable of fulfilling the purpose it was designed for.  
The cost function $\cost(n)$ is intended to provide a means to quantitatively express the efforts required by an attacker to compromise a given node $n$. 
We realise that the instantiation of this cost function may be difficult in some cases. 
However, there have been many remarkable advances in this direction over the last years. 
For example, the \textit{Common Vulnerability Scoring System (CVSS)} is a standard security effort that provides a means to quantify software vulnerabilities in the form of a single numerical score \cite{CVSS}.  
This score reflects the severity of a vulnerability and considers many aspects such as complexity, exploitability, impact, among others. 
Overall, security languages and frameworks such as CVSS underpin automation and promote knowledge exchange that can be used to assess the security of industrial networks. 

\begin{algorithm}[!t]
	\DontPrintSemicolon
	\SetLine
	\Name{$\rem(G, X)$}
	\KwIn{Graph $G=(V,E)$, Nodes to remove $X$}
	\KwOut{Updated graph $G=(V,E)$}
	
	\BlankLine  	
	\While{$X$ is not empty}{
		Node $n \gets X.pop()$\tcp*{get first node} 	
		
		$M \gets \{x \in V: (n,x) \in E\} $\tcp*{nodes reached by $n$} 	
		
		\For{$x \in M$}{
			\If{$(x \in V_{AT})$ or $(x \in \andnodes)$ or $(x \in \ornodes \land \degin(x)=1)$ }{					
				$X.append(x)$\tcp*{$x$ must be removed}		
			}			
		}
		
		$V = V - \{n\}$\tcp*{remove $n$ from $G$} 			
		$E = E - \{(v,w) \in E : v=n \lor w=n\}$\tcp*{rem. edges} 	 					
	}
	
	\Return $G = (V,E)$
	
	\caption{Logical node removal $\rem(G, X)$}	
	\label{alg:removeNode}
\end{algorithm}

Since CPS components (graph nodes) logically depend on others to work properly, our adversarial model considers that the compromise of a node will also affect the operation of the nodes that depend on it. Therefore, such impact is passed on to other nodes following a logic-style propagation. 
Algorithm~\ref{alg:removeNode} describes this process from a node removal standpoint in the form of a function $\rem(G, X)$. 
The function takes as input an AND/OR graph $G$ and a set of nodes to remove $X$, and returns $G$ after deleting the nodes in $X$ as well as the nodes that logically depend on them, and every edge related to the removed nodes. 
Essentially, for each node $n$ that must be removed, Algorithm \ref{alg:removeNode} analyses the nodes that depend on $n$ (set $M$, line 3). 
A node $x$ that depends on $n$ will be affected only if $x$ is an atomic node, an AND node, or an OR node with only one input left (line 5). 
In any of these cases, node $x$ is queued for removal (line 6). 
In other words, only an OR node with more than one input will remain in the graph when one of its inputs is removed. 
Finally, each node marked for removal is deleted from $G$ along with its respective edges (lines 9-10). 
In what follows, we use $\rem(G, X)$ to express the impact of compromised nodes as well as the resulting graph when such nodes are removed from the network.

\subsection{Simple example}
\label{sec:simple_example}
\begin{figure}[!h]
	\centering
    \includegraphics[scale=0.28]{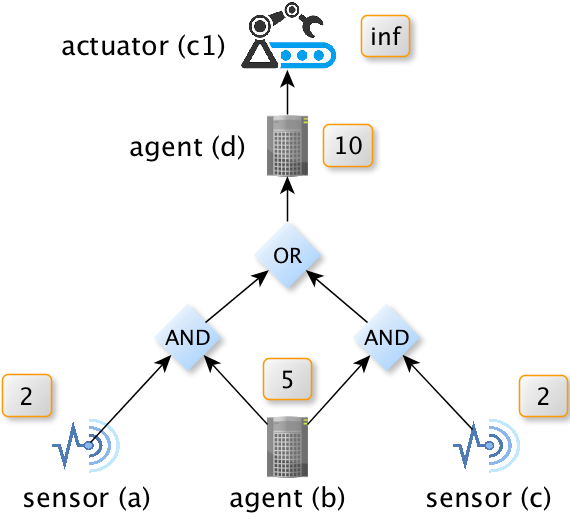}	
	\caption{AND/OR graph with sensors, software agents and actuators}
	\label{fig:simple-example1}
\end{figure}

Let us consider a simple example involving the scenario illustrated in Figure \ref{fig:simple-example1}. 
This setup involves five CPS components in the form of atomic nodes: 2 sensors ($a$ and $c$), 2 software agents ($b$ and $d$), and 1 actuator ($c1$). 
In addition. the graph includes 2 AND nodes and 1 OR node that express the logical dependencies among the CPS components. 
Each CPS component also has an associated value that represents its compromise cost where \textit{inf} means infinite.  

The graph reads as follows: the actuator $c1$ depends on the output of software agent $d$. Agent $d$ in turn has two alternatives to work properly; it can use either the readings of sensor $a$ and the output from agent $b$ together, or the output from agent $b$ and the readings of sensor $c$ together. 
Now, considering the compromise cost of each CPS component, the question we are trying to answer is: which nodes should be compromised in order to disrupt the operation of actuator $c1$, with minimal effort (cost) for the attacker? In other words, what is the least-effort attack strategy to disable actuator $c1$?

Our example involves many attack alternatives, however, only one is minimal. For example, the attacker could compromise node $d$ and thus, the target node $c1$ would be successfully disconnected from the graph. However, this strategy has cost 10. Another option would be to compromise node $b$, with a lower cost of 5. Because node $b$ feeds both AND nodes, these will be disrupted and consecutively the OR node, which in turn will affect node $d$ and finally node $c1$. 
In terms of costs, however, the optimal strategy for the attacker in this case is to compromise nodes $a$ and $c$ with a total cost of $4$. 
From a defence perspective, we understand this minimal cost as a metric that represents the security level of the system we are trying to protect. 
In the next section, we formalise our security metric in the form of an optimisation problem and explain the resolution approach. 

\section{Security metric}
\label{sec:metric}

\subsection{Problem definition}
Let $W$ be an industrial network, $G = (V,E)$ an AND/OR graph representing the operational dependencies in $W$, and $t$ a target network node. The objective of our security metric, $\mu(G,t)$, is to identify the set of nodes $X=\{m_1, \ldots, m_h\}$ that must be compromised in order to disrupt the normal operation of target node $t$, with minimal cost for the attacker. 

More formally, $\mu : \gdomain \times V \rightarrow 2^{V}$ is defined as follows: 

\begin{equation}
\begin{array}{c}
\displaystyle \mu(G,t) = \underset{X \subseteq V_{AT}}{\text{argmin} } \Big( \sum_{m_i \in X} \cost(m_i) \Big) \\  
\textrm{s.t.}\\
\displaystyle \wcc(\rem(G, X)) \geq 2 \lor X = \{t\}\\
\end{array}
\end{equation}

where the solution with minimal cost must be either node $t$ or a set of nodes $X$ such that, if removed, $t$ gets disconnected from the graph. This is achieved with function $\wcc(G)$, which computes the number of weakly connected components in $G$, that is, the number of connected components when the orientation of edges in $G$ is ignored. In other words, the restriction on $\wcc(G)$ ensures that the target node $t$ is disconnected from a non-empty set of nodes on which $t$ depends (directly or indirectly) to function properly. 

We also define $\kappa : \gdomain \times V \rightarrow R_{\geq0}$, a numerical version of the metric that is based on the previous formulation as follows: 

\vspace{-0.2cm}
\begin{equation}
\begin{array}{c}
\displaystyle \kappa(G,t) = \sum_{m_i \in \mu(G,t)} \cost(m_i)\\
\end{array}
\end{equation}

It is important to note that if we assign the same unit cost to every node (i.e. $\cost(v) = 1, \forall v \in V_{AT}$), the metric will indicate the minimum number of nodes that must be compromised in order to bring down the system. Alike, though from a fault-diagnosis perspective, the metric will determine the minimum amount of components that must concurrently fail in order to make the system non-operational. 

From a graph-theoretical perspective, our security metric looks for a minimal weighted vertex cut in AND/OR graphs. As mentioned in Section \ref{sec:rw}, this is an NP-complete problem \cite{Desmedt2002, Desmedt2004, DesmedtUsingFF, Souza2013}. However, a proper transformation of the problem can leverage advanced mechanisms from other fields and take advantage of very efficient techniques for this sort of problem. In this paper, we address our problem from a logical perspective, and more precisely, from a satisfiability point of view.

\subsection{Metric resolution approach via graph transformation} 
Given a target node $t$, the input graph $G$ can be used as a map to decode the dependencies that node $t$ relies on. Since these dependencies are presented as a logical combination of components connected with AND and OR operators, we say that node $t$ is fulfilled (or can operate normally) if the logical combination is satisfied. In turn, these dependencies may also have previous dependencies, and therefore, they must be also satisfied. In that sense, $G$ can be traversed backwards in order to produce a propositional formula that represents the different ways in which node $t$ can be fulfilled. We call this transformation $\form(t)$. 
To illustrate this idea, let us consider the previous example scenario (presented in Section \ref{sec:model}). 
In this case, $\form(c1)$ returns the following formula: 
\vspace{-0.05cm}
\begin{lstlisting}[
label={lst:form-step1}, 
mathescape=true, language=bash, %caption=Security metric resolution, 
basicstyle=\martinListingFontsize\sffamily, 
keywordstyle=\bfseries\color{green!40!black}, 
commentstyle=\itshape\color{purple!40!black},
identifierstyle=\color{blue},
stringstyle=\color{orange}, 
frame=single, 
%numbers=left,
numberstyle=\tiny\color{gray},
stepnumber=1, 
belowcaptionskip=5em,
belowskip=3em, 
captionpos=b, 
%xleftmargin=.125\textwidth, xrightmargin=.125\textwidth
]
                $c1 \land ( d \land ( ( a \land b ) \lor ( b \land c ) ) )$
\end{lstlisting}
\vspace{-0.5cm}

\vspace{-0.2cm}
The goal of the attacker, however, is precisely the opposite, i.e., to disrupt node $t$ somewhere along the graph. 
Therefore, we are actually interested in satisfying $\neg \form(t)$, which describes the means to disable $t$, as follows: 
\begin{lstlisting}[
label={lst:form-step2}, 
mathescape=true, language=bash, %caption=Security metric resolution, 
basicstyle=\martinListingFontsize\sffamily, 
keywordstyle=\bfseries\color{green!40!black}, 
commentstyle=\itshape\color{purple!40!black},
identifierstyle=\color{blue},
stringstyle=\color{orange}, 
frame=single, 
%numbers=left,
numberstyle=\tiny\color{gray},
stepnumber=1, 
belowcaptionskip=5em,
belowskip=3em, 
captionpos=b, 
%xleftmargin=.125\textwidth, xrightmargin=.125\textwidth
]                   
               $\neg ( c1 \land ( d \land ( ( a \land b ) \lor ( b \land c ) ) ) )$
\end{lstlisting}
\vspace{-0.5cm}

\vspace{-0.2cm}
Under that perspective, a logical assignment such that $\neg \form(t) = true$ will indicate which nodes must be compromised (i.e. logically falsified) in order to disrupt the operation of the system. Finding such an assignment constitutes a Satisfiability (SAT) problem \cite{Cook1971}. A SAT problem essentially looks for an assignment of truth values to the variables of a logical formula such that the formula evaluates to $true$. Normally, SAT formulations consider the input formula in conjunctive normal form (CNF). Converting an arbitrary boolean formula to CNF can be naively tackled by using De Morgan and distributive laws, which leads to the following formula: 

\begin{lstlisting}[
label={lst:form-step3}, 
mathescape=true, language=bash, %caption=Security metric resolution, 
basicstyle=\martinListingFontsize\sffamily, 
keywordstyle=\bfseries\color{green!40!black}, 
commentstyle=\itshape\color{purple!40!black},
identifierstyle=\color{blue},
stringstyle=\color{orange}, 
frame=single, 
%numbers=left,
numberstyle=\tiny\color{gray},
stepnumber=1, 
belowcaptionskip=5em,
belowskip=3em, 
captionpos=b, 
%xleftmargin=.125\textwidth, xrightmargin=.125\textwidth
]                   
         $(\neg c1 \lor \neg d \lor \neg a \lor \neg b) \land (\neg c1 \lor \neg d \lor \neg b \lor \neg c )$
\end{lstlisting}
\vspace{-0.5cm}

\vspace{-0.2cm}

However, such an approach might lead to exponential computation times over large graphs, thus only being able to scale up to a few hundred nodes. To avoid this issue, we use the Tseitin transformation \cite{Tseitin70}, which essentially produces a new formula in CNF that is not strictly equivalent to the original formula (because there are new variables) but is equisatisfiable. This means that, given an assignment of truth values, the new formula is satisfied if and only if the original formula is also satisfied. An example of how the Tseitin transformation works can be found in Appendix \ref{sec:annex-tseitin-example}. Since the Tseitin transformation adds new variables during the process, the new formula is larger in size than the original one (we omit the transformed formula for our example since it has 15 variables and 27 clauses). However, the Tseitin transformation can be done in polynomial time, as opposed to the naive CNF conversion approach that can ramp up to exponential computation times in the worst case.

\subsection{Satisfiability formulation} 
\label{sec:sat-formulation}
When a CNF formula also involves weights on each clause, the problem is called MAX-SAT \cite{Davies2011}. A MAX-SAT problem consists in finding a truth assignment that maximises the weight of the satisfied clauses. Equivalently, MAX-SAT minimises the weight of the clauses it falsifies \cite{Davies2011}. When a set of clauses must be forcibly satisfied (called \textit{hard clauses}), the problem is denominated Partial MAX-SAT and it works on a subset of clauses (denominated \textit{soft clauses}) that can be falsified if necessary. 
If the \textit{soft clauses} have non-unit weights, the problem is called Weighted Partial MAX-SAT and it will try to minimise the penalty induced by falsified weighted variables. We use the latter to address our problem. Reconsidering our example scenario, the hard clauses are those involved in the CNF formula as follows:
\vspace{-0.2cm}
{\renewcommand{\arraystretch}{1.15}
\begin{table}[!h]
	\centering
	\begin{tabular}{|c|}
		\hline
		$\neg c1 \lor \neg d \lor \neg a \lor \neg b$\\ 
		\hline
		$\neg c1 \lor \neg d \lor \neg b \lor \neg c$\\
		\hline
	\end{tabular}
\end{table}
}
\vspace{-0.35cm}

\hspace{-0.34cm}whereas soft clauses correspond to each atomic node in the graph with their corresponding penalties (costs) as follows: 
\vspace{-0.3cm}
{\renewcommand{\arraystretch}{1.15}
\begin{table}[!h]
	\centering
	\begin{tabular}{|c|c|c|c|c|}
		\hline
		{$a$} & {$b$} & {$c$} & {$d$} & {$c1$} \\
		\hline
		{$\cost(a)=2$} & {$\cost(b)=5$} & {$\cost(c)=2$} & {$\cost(d)=10$} & {$\cost(c1)=inf$} \\
		\hline
	\end{tabular}	
\end{table}
}

Therefore, a MAX-SAT solver will try to minimise the number of falsified variables as well as their weights, which in our problem equals to minimise the compromise cost for the attacker. As shown in Section \ref{sec:analytical-exp}, current SAT solvers are able to handle this family of problems at a very decent large scale (dozens of thousands of variables), and they usually involve state-of-the-art techniques to tackle satisfiability problems, pseudo-boolean problems and optimisation procedures \cite{Berre2010}. The following section details our strategy to compute the security metric. 

\section{Computation strategy}
\label{sec:computation-strategy}

\subsection{Logical transformation}
\label{sec:logical_transformation}

\begin{algorithm}[!t]
    \DontPrintSemicolon
    \SetKwFunction{getSentence}{getSentence}
    \SetLine
    \Name{$getSentence$}
    \Global{Graph $G=(V,E)$}
    \KwIn{Node $n$, Visited nodes $M$}
    \KwOut{Logical sentence $p$}
    
    \BlankLine  
    $M' \gets M \cup \{n\}$\tcp*{mark $n$ as visited} 	
    
    \If(\tcp*[f]{node $n$ is atomic}){$n \in V_{AT}$}{	
        $x \gets incomingNode(G, n)$\tcp*{predecessor} 				
        \uIf(\tcp*[f]{null $x$ or visited $x$}){$not(x) \,||\, x \in M$}{
            $p \gets$ $n$ \tcp*{atomic sentence} 		
        }\Else{				
            $s \gets getSentence(x, M')$\tcp*{recursive call} 	
            $p \gets$ $($ $\concat$ $n$ $\concat$ $\land$ $\concat$ $s$ $\concat$ $)$\tcp*{concat with $\concat$} 
        }
    }
    
    $X \gets incomingNodes(G, n)$\tcp*{nodes reaching $n$} 		
    $X \gets filter(X, visited)$\tcp*{unseen nodes only} 		
    
    \If(\tcp*[f]{$n$ has AND type}){$n \in \andnodes$}{					
        $p \gets getMultiSentence(X, \land, M')$\tcp*{AND operator}		
    }
    
    \If(\tcp*[f]{$n$ has OR type}){$n \in \ornodes$}{			
        $p \gets getMultiSentence(X, \lor, M')$\tcp*{OR operator} 				
    }
    
    \Return p
    
    \caption{Main logical sentence builder (recursive)}	    
    \label{alg:getSentence}
\end{algorithm}

Given a directed AND/OR graph $G=(V,E)$ and a target node $t \in V_{AT}$, we first produce a propositional formula that represents the logical semantics of $G$ with regards to $t$, i.e. the logical conditions that must be satisfied to fulfil $t$. We denote this transformation as $\form(t)$, which is described in Algorithm \ref{alg:getSentence}. 
The formulation process starts at $t$ and traverses $G$ backwards, expanding logical conditions as needed, until nodes with no incoming edges are reached. 

Algorithm \ref{alg:getSentence} moves recursively through the graph and builds a valid logical sentence considering three main cases that depend on the type of node being analysed. 
Atomic nodes ($n \in V_{AT}$) constitute the first case, which is expanded recursively if node $n$ has a predecessor. 
Atomic nodes only have one incoming edge by definition, with the exception of the source that has none. The other two cases correspond to AND/OR nodes respectively, and are treated in a similar way. In these cases, the algorithm calls a second function, described in Algorithm \ref{alg:getMultiSentence}, which essentially builds sub-sentences for each predecessor of the AND/OR node (stored in NodeList $X$) and joins them using the appropriate operator $op \in \{\land,\lor\}$. 

There are two important aspects about the logical transformation that are worth to mention. 

\textbf{1. The $\land$ operator on atomic recursive calls}. 
In line 8, Algorithm \ref{alg:getSentence} builds a sentence with the current node ($n$) and the sentence obtained from its predecessor ($x$), by using the AND operator ($\land$). 
The reason for using the AND operator relies on the semantics of the graph $G$, which represents dependencies between components. 
Because $n$ depends on its predecessor $x$, node $n$ can only be fulfilled if its predecessor $x$ is fulfilled ($x$ can be an AND/OR node as well), and therefore, we state so using the $\land$ operator. 

\begin{algorithm}[!t]
	\DontPrintSemicolon
	\SetLine
	\Name{$getMultiSentence$}
	\KwIn{NodeList $X$, Operator $op$, Visited nodes $M$}
	\KwOut{Logical sentence $p$}
	\BlankLine  
	
	\If(\tcp*[f]{empty set of nodes}){$X = \{\}$}{			
		\Return $true$	
	}
	$p \gets ($\tcp*{open sentence} 		
	
	\For{$i = 0;\ i < |X|-1;\ i = i + 1$}{
		$x \gets X.get(i)$\tcp*{get node from list} 					
		$s \gets getSentence(x, M)$\tcp*{compute sub-sentence} 								
		$p \gets p$ $\concat$ $s$ $\concat$ $op$ \tcp*{concat with $\concat$} 					
	}
	$x \gets X.get(|X|-1)$\tcp*{get last node} 					
	$s \gets getSentence(x, M)$\tcp*{compute sub-sentence} 						
	$p \gets p$ $\concat$ $s$ $\concat$ $)$ \tcp*{close sentence} 				
	\Return $p$
	\caption{Logical multi-sentence builder ($\land$,$\lor$)}	
	\label{alg:getMultiSentence}
\end{algorithm}

\textbf{2. Cycles}. 
Normally, AND/OR graphs are acyclic \cite{Souza2013}. 
However, the meaning of cycles in AND/OR graphs representing dependencies might be debatable. 
In this work, we aim at tackling the general case where the input graph $G=(V,E)$ may also contain cycles. 
Our approach to deal with cycles is to keep a record of the nodes that have been analysed so far. 
In that sense, Algorithm \ref{alg:getSentence} controls cycles by using a set of visited nodes. 
Nodes that have already been visited are not expanded again in deeper explorations coming from them. 
This is possible because the truth value of an already visited node is present in an earlier part of the formula and connected to their predecessors via the $\land$ operator. 
For the atomic case, this is directly implemented in line 5, while for complex AND/OR cases, visited predecessors are previously filtered in line 12. 
Annex \ref{sec:annex-cycles} illustrates an example of the cycle handling process.

\subsection{Weighted Partial MAX-SAT approach} 
\label{sec:sat-resolution}
Given an input graph $G=(V,E)$ and a target node $t$, we model our security metric  $\mu(G,t)$ as a Weighted Partial MAX-SAT problem where the weights are provided by the cost function $\cost(v)$ for each node $v \in V_{AT}$. 
Our methodology includes the following steps: 

\begin{enumerate}[1.]
	\item We first transform the dependency graph $G$ into an equivalent logical representation, $\form(t)$, as described in Section \ref{sec:logical_transformation}.
	\item Then, we normalise the input formula that represents the objective of the attacker using the Tseitin transformation, i.e., $h(v) = CNF(\neg \form(v))$, where
	$h(v) = (v_{1i} \lor \ldots \lor v_{1j}) \land \ldots \land (v_{hi} \lor \ldots \lor v_{hj}) $. 	
	\item We then define the clauses of $h(v)$ as hard clauses; that is, we force each clause $(v_{hi} \lor \ldots \lor v_{hj})$ to evaluate to $true$.  
	\item Finally, we define a soft clause for each atomic node in the graph, and assign its corresponding weight as \hbox{$(v, \cost(v)), v \in V_{AT}$}. 
\end{enumerate}

The last step tells the MAX-SAT solver that each soft clause (each node $v$ of the graph) can be falsified with a certain penalty $\cost(v)$, which is the cost required for the attacker to compromise $v$. 
Since the solver tries to minimise the total weight of falsified variables, a solution to this problem yields a minimum vertex cut in the graph that models our CPS system. 
Therefore, the overall process provides a set of critical components with minimal cost for the attacker such that, if removed, it would render the cyber-physical system into a non-operational state. 
In the following section, we present the technical details of our implementation prototype and explore our initial example illustrating the overall process step by step. 

\section{ \tool $-$ Implementation prototype}
\label{sec:impl}

\begin{figure}[!t]
    \centering
    \includegraphics[scale=0.132]{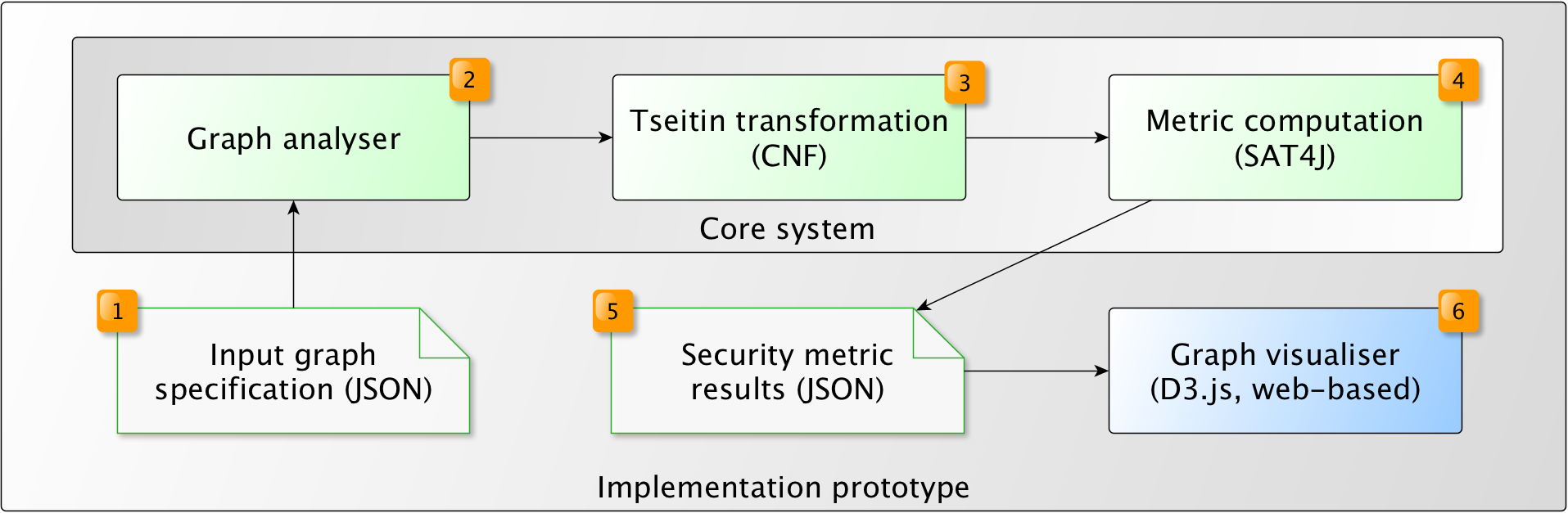}
    \caption{\tool overall architecture}
    \label{fig:architecture}
\end{figure}

In this section, we describe our implementation prototype called \tool (Metric Analyser for Industrial Control Systems), pronounced as \textit{metaphorics}. The overall architecture of \tool is illustrated in Figure \ref{fig:architecture}, and it involves two main modules:   
\begin{inparaenum}[(i)]
    \item a core system in charge of analysing the input graph and computing the security metric, and
    \item a Web-based visualisation system that displays the graph as well as the critical nodes indicated by the security metric. 
\end{inparaenum} 

\begin{figure}[h]
    \centering
    \includegraphics[scale=0.45]{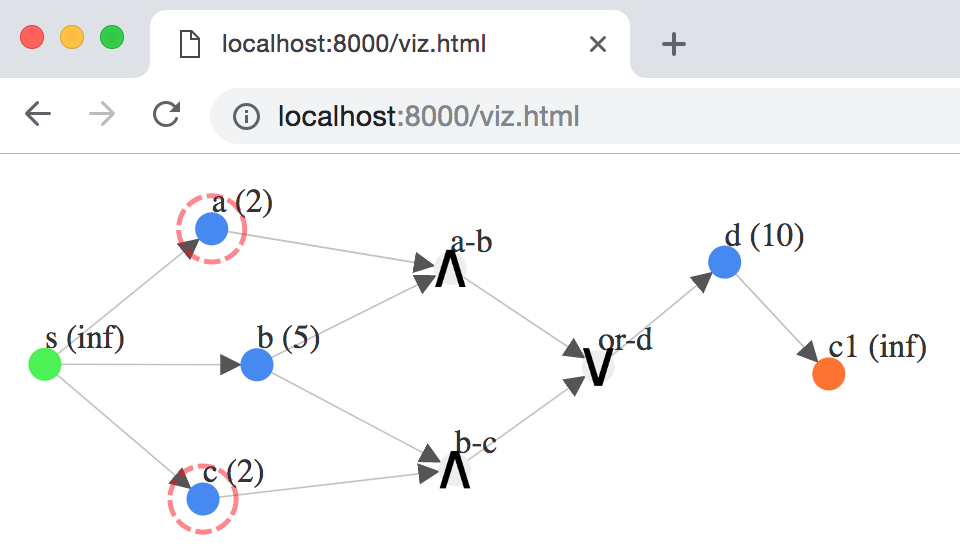}
    \caption{Graph viewer - metric resolution}
    \label{fig:example1-viewer}
\end{figure}

The core system has been purely developed in Java and it can be executed as a single runnable JAR. Initially (step~1), the tool consumes an input graph represented in JSON (JavaScript Object Notation). If the input graph has more than one node with no incoming edges, the graph analyser creates an artificial source node, denoted as $s$, which is linked to each one of these nodes in order to produce a single-source graph (step~2). This is done to simplify graph-processing algorithms.
Within our simple example (see Section \ref{sec:model}), nodes $a$, $b$, and $c$ have no incoming edges, and therefore, the following edges are also added: $(s,a)$, $(s,b)$, and $(s,c)$. This can be observed in Figure~\ref{fig:example1-viewer} (later described). 

Afterwards, our prototype implements the Tseitin transformation (step~3) to convert an arbitrary boolean formula into an equisatisfiable CNF formula \cite{Tseitin70} in order to provide efficient computation times. For the MAX-SAT resolution process (step~4), we use SAT4J, a Java-based library for solving boolean satisfaction and optimisation problems \cite{SAT4J, Berre2010}. In the case of ties (even cost for two or more solutions), the tool selects the solution with the minimum number of nodes. The outcome of this process is also represented in JSON format (step~5) and is used to feed the second main component of our tool (step~6). 
The latter is an interactive graph visualiser, built on top of the D3.js technology \cite{D3.js}, whose objective is to provide visual means to understand dependencies among nodes and manipulate critical nodes. Figure \ref{fig:example1-viewer} shows the metric resolution for our example scenario. The tool displays critical nodes surrounded by dashed red circles and allows the user to validate the solution by interactively removing them until the target is disabled.  

Annex \ref{sec:tool-execution} provides a full description of the execution of \tool over our example scenario, including technical details about input and output JSON formats. 
\tool is publicly available at \cite{BarrereMeta4icsGithub}. 
In the next section, we present an experimental evaluation of our approach where we analyse performance and scalability aspects.

\section{Experimental evaluation}
\label{sec:analytical-exp}

In order to evaluate the feasibility of our approach, we have conducted an extensive set of experiments based on synthetic pseudo-random AND/OR graphs of different size and composition. These experiments have been performed using a standard MacBook Pro (Mid 2014), 
2.8 GHz Intel Core i7 processor, 16 GB of memory. The construction procedure for an AND/OR graph of size $n$ is as follows. 
We first create the target node. Afterwards, we create a predecessor which has one of the three types (atomic, AND, OR) according to a probability given by a compositional configuration predefined for the experiment. For example, a configuration of $(60,20,20)$ means 60\% of atomic nodes, 20\% of AND nodes and 20\% of OR nodes. We repeat this process creating children on the respective nodes until we approximate the desired size of the graph, $n$. 

\begin{figure}[!b]
	\centering
	\includegraphics[scale=0.7]{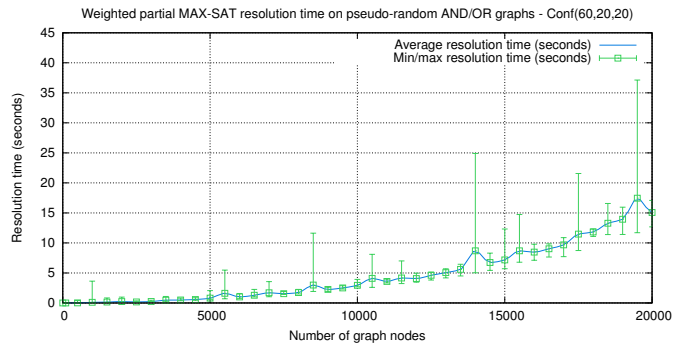}
    \caption{Scalability and performance evaluation up to 20000 nodes}
	\label{fig:exp1}
\end{figure}

Figure \ref{fig:exp1} shows the behaviour of our methodology when the size of the input graph increases. 
In this experiment, we produce pseudo-random AND/OR graphs of size $n$ and a compositional configuration of (60,20,20). 
The size $n$ varies as $n \in [0, 500, 1000, 1500, \ldots, 20000]$, and we iterate the evaluation process 10 times for each value of $n$. 
The solid line shows the average values obtained for graphs of size $n$ while the vertical bars indicate shortest and largest computation times for each value of $n$. As we explore later in this section, the structure of the logical formulation varies according to the structure of the input graph (e.g. more ANDs than ORs), and therefore, the time required by the CNF converter and SAT solver might vary as well, which explains the vertical bars. In the general case, however, we have observed very good results in terms of performance and scalability. 
For example, for graphs with 10000 nodes, the average resolution time is about 3 seconds, while for graphs with 20000 nodes, the average time is around 15 seconds. Note that scalability here is understood from a computational standpoint rather than a control systems perspective. 

\begin{figure}[t]
    \centering
    \includegraphics[scale=0.7]{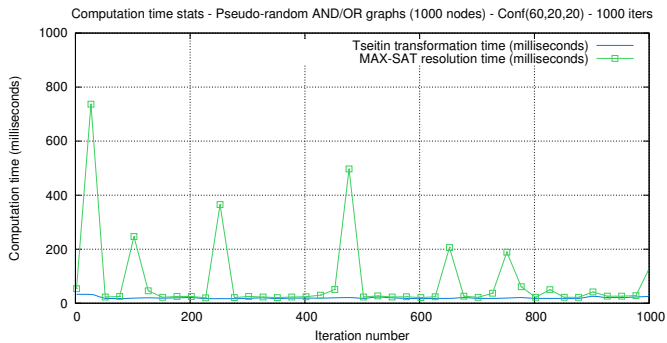}
    \caption{Performance evaluation (1000 nodes x 1000 iterations)}
    \label{fig:exp2}
\end{figure}

In order to analyse the variability observed in computation times due to the structure of the formulas, we have taken a closer look at the two main processes that govern the overall behaviour of the strategy: the Tseitin transformation and the MAX-SAT resolution. 
Figure \ref{fig:exp2} shows the results of a 1000-iteration experiment using AND/OR graphs with 1000 nodes and a (60,20,20) configuration. 
In general, we have observed that while the Tseitin transformation time is stable across all iterations, the MAX-SAT resolution process requires more time to solve the problem in some graphs than others. This happens because some graphs induce formulas involving longer sequences of AND or OR operators connecting with different combinations of graph nodes, which incurs in variable computation times. 

\begin{figure}[b]
    \centering
    \includegraphics[scale=0.7]{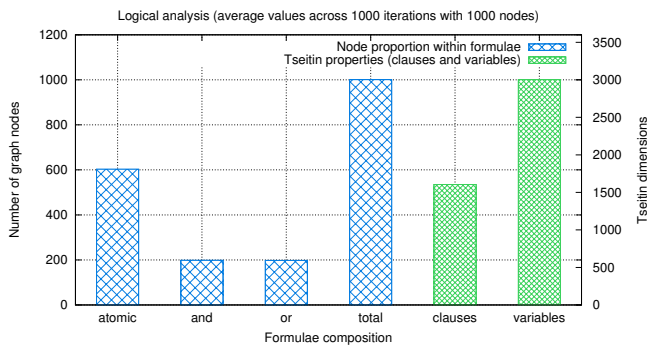}
    \caption{Logical compositional analysis (1000 nodes x 1000 iterations)}
    \label{fig:exp3}
\end{figure}

We have also observed that the number of clauses and variables within the transformed Tseitin formulas involve similar patterns for this and other configurations. For example, Figure \ref{fig:exp3} details the formulae composition for a (60,20,20) configuration. 

\begin{figure}[t]
    \centering
    \includegraphics[scale=0.72]{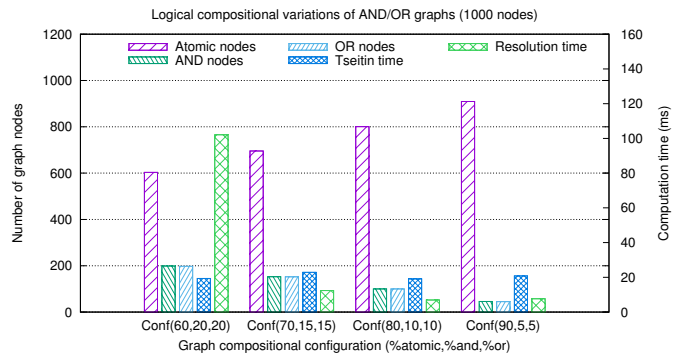}
    \caption{Comparative analysis of different logical compositions (1000 nodes)}
    \label{fig:exp4}
\end{figure}

In average, the Tseitin formulas involve 1500 clauses and 3000 variables for graphs with 1000 nodes and the (60,20,20) configuration. 
This means that each clause generally involves the disjunction of two or three variables. In order to better understand how the complexity of the formulas may impact the overall strategy, we have also experimented with different composition configurations for graphs with 1000 nodes. 
Figure \ref{fig:exp4} shows the obtained results on four different configurations including the previous (60,20,20) distribution. 

\begin{figure}[b]
    \centering
    \includegraphics[scale=0.7]{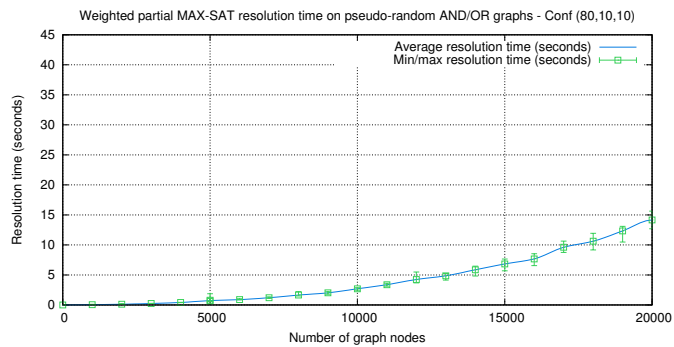}
    \caption{Scalability and performance evaluation - Conf(80,10,10)}
    \label{fig:exp5}
\end{figure}

While the Tseitin transformation shows almost a constant behaviour, we can observe a dramatic reduction in the average MAX-SAT resolution time as the number of AND/OR nodes decreases. This phenomenon occurs because the graphs now involve more dependent nodes in sequence with less AND/OR nodes among them. In addition, OR nodes have a higher impact in the resolution time since any fulfilled input may enable this connector, while AND nodes only require one disconnected input to be disabled. In order to confirm these observations, we have conducted the same scalability experiment up to 20K nodes, but now with a different configuration using a (80,10,10) distribution. The results are shown in Figure~\ref{fig:exp5}. As expected, we can observe that the variability of the experiments (vertical green bars) is now much lower since the structure of the formulas pose less restrictions to find the optimal solutions. 

Overall, the obtained results indicate that our methodology can efficiently scale to large graphs involving thousands of nodes and compute the proposed security metric in a matter of seconds. Moreover, considering that many CPS scenarios and their respective graph models might not be that large, and assuming that an automated mechanism to produce such models is in place, our approach could be used to perform security assessments upon environment changes in \hbox{nearly real-time}.

\textbf{Discussion}. 
Scalability is an essential aspect when dealing with evolving and growing environments. However, it is also certain that ICS networks have been designed with some underlying structure in mind, i.e., they have not been created chaotically. 
In that sense, they usually present some organisational characteristics that we can leverage to address complexity even more. 
We argue that insightful structural information about the network (e.g. clusters, zones, regions, subnets), may be used to reduce large graphs into smaller subproblems and compose their solutions. 
However, understanding and generalising structural properties of real-world industrial settings is a challenging goal. 
For example, while oil and gas facilities may involve kilometres of pipes and sensors depicting elongated and thin graphs with clear articulation points, other industrial settings may be translated into more dense graphs with highly interconnected components. 
Additional research questions such as what numbers are representative regarding size and classes of components (hundreds, thousands, tens of thousands?) are also important. In that context, one research direction already scheduled as future work is to further study the structural properties of ICS networks that may help improve the analysis of the graphs that represent them. 

In addition, time variations, which are closely related to the compositional configuration of the graphs and their size, are also important. However, characterising them properly is a challenging task. While small logical formulas that can be analysed by hand may present small variations in the order of nanoseconds in a multi-task operating system (thus hard to analyse), larger graphs with more evident variations in the order of seconds or milliseconds may usually involve hundreds or thousands of nodes. In that context, part of our future work involves a deeper analysis within the internal MAX-SAT solving strategies in order to better characterise these variations and address them more efficiently.  

In the next section, we present a thorough case study based on a realistic water transport network where we analyse and discuss the use and applications of our security metric. 

\newcommand{\nn}[1]{{\color{black} #1}}

\section{Case study on Water Transport Networks and Extended Security Applications}
\label{sec:case_study}
 
Security metrics are useful in many contexts and applications, especially in the case of critical infrastructure systems which rely on industrial control systems for their safe and efficient operation. Our case study is focused on water transport networks (WTNs). 
In these systems, operators need to understand how exposed their systems are to attacks, and they should be able to evaluate different ``what-if'' scenarios to improve their security posture. These scenarios enable water utility operators to understand how configuration changes may increase or decrease the overall security of the system. 

As mentioned before, our approach considers as input an AND-OR graph that models the system under analysis in order to quantify its security level, that is, to inform operators about how secure their system is. In this paper, we do not focus on the generation of AND-OR graphs, however, we do provide in this section some generation insights to model water transport networks, even though this typically depends on the nature of the plant being analysed. In this case study, we are focused on a realistic water transport network, structurally similar to the one illustrated earlier in Figure \ref{fig:large-scale-wdn}, Section \ref{sec:intro}.  
We have observed that, for this particular type of industrial control system, there exist basic structural constructs that appear repeatedly, thus forming patterns in large setups. In that context, and for the sake of clarity, we first present in this section a description of these small constructs that form the basis of large scenarios, we explain how these simple models can scale to larger ones, and then we provide a thorough analysis of our approach using compact scenarios. 
Finally, we present further uses of our security metric as well as other potential applications.

\subsection{Case study description}

\begin{figure}[t]
	\centering
    \includegraphics[scale=0.25]{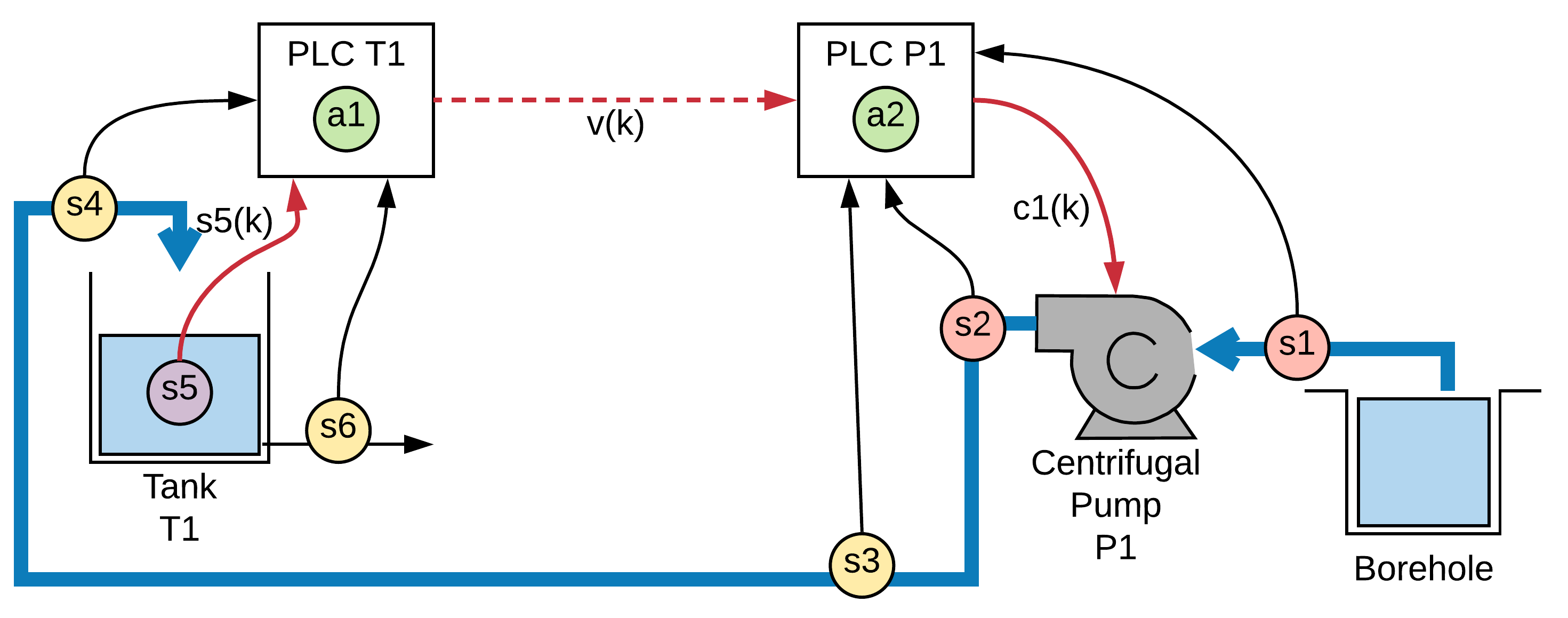}
	\caption{Basic water transport network component \cite{Nicolau2018}} 
    \vspace{-0.5cm}
	\label{fig:case-study-1}
\end{figure}

Figure \ref{fig:case-study-1} depicts a realistic basic structural component of a water transport system, which constitutes a typical configuration of several water utilities as confirmed by water utility experts and other scientific sources~\cite{Trifunovic2006}. 
This basic component may also appear multiple times in a real water transport network, as later described in this section. 
In this case study, drinking water is extracted from a water source (e.g. a borehole or another tank) using a pump. The pump increases the water pressure which pushes the water into a tank, which may be located a few kilometres away at a higher elevation. The water tank is then used to provide water to consumers, as well as to transfer water to other subsystems. 
Additional details of this scenario can be found in \cite{Nicolau2018}.

We consider the following sensing elements which appear in the topology of the case study: a pressure sensor before the pump ($s1$), a pressure sensor after the pump ($s2$), and a water flow sensor ($s3$) measuring the pump outflow. At the water tank, flow sensors ($s4$, $s6$) are also installed for monitoring the inflow and outflow respectively. 

For its operation, the control system is comprised of Programmable Logic Controllers (PLCs) connected to the aforementioned sensors and actuators. In this example, control is achieved with two PLCs, with the one situated at the pump and the other at the water tank. These PLCs are connected to the system's sensors and actuators, and execute programs to achieve the control objectives. More specifically, the sensing node $s5$ provides the water level state measurement $s5(k)$ to the agent $a1$ in PLC-T1, where $k$ is the discrete time step. Then, the control logic is executed, and the result $v(k)$ is transmitted to PLC-P1, where another control logic $a2$ is executed. This control logic instructs the contactor (i.e., an electrically operated relay) through a signal $c1(k)$ to turn on/off the pump, should the pump flow $s3$ be below a threshold. 

Typically, these systems are set up using the minimum configuration. In our example scenario, measurements from sensors $s5$ and $s3$ are necessary in order to operate the system. As a result, if there is a problem (or an attack) affecting one of the two sensors, the system will not be able to function properly. From a water operator point of view, the challenge would be to measure how secure the current configuration of the system is, and how security can be enhanced by adding new elements (such as sensors and software agents) into the system. 
We discuss redundancy aspects in Section \ref{sec:counter}.  

\begin{figure}[t]
	\centering
    \includegraphics[scale=0.20]{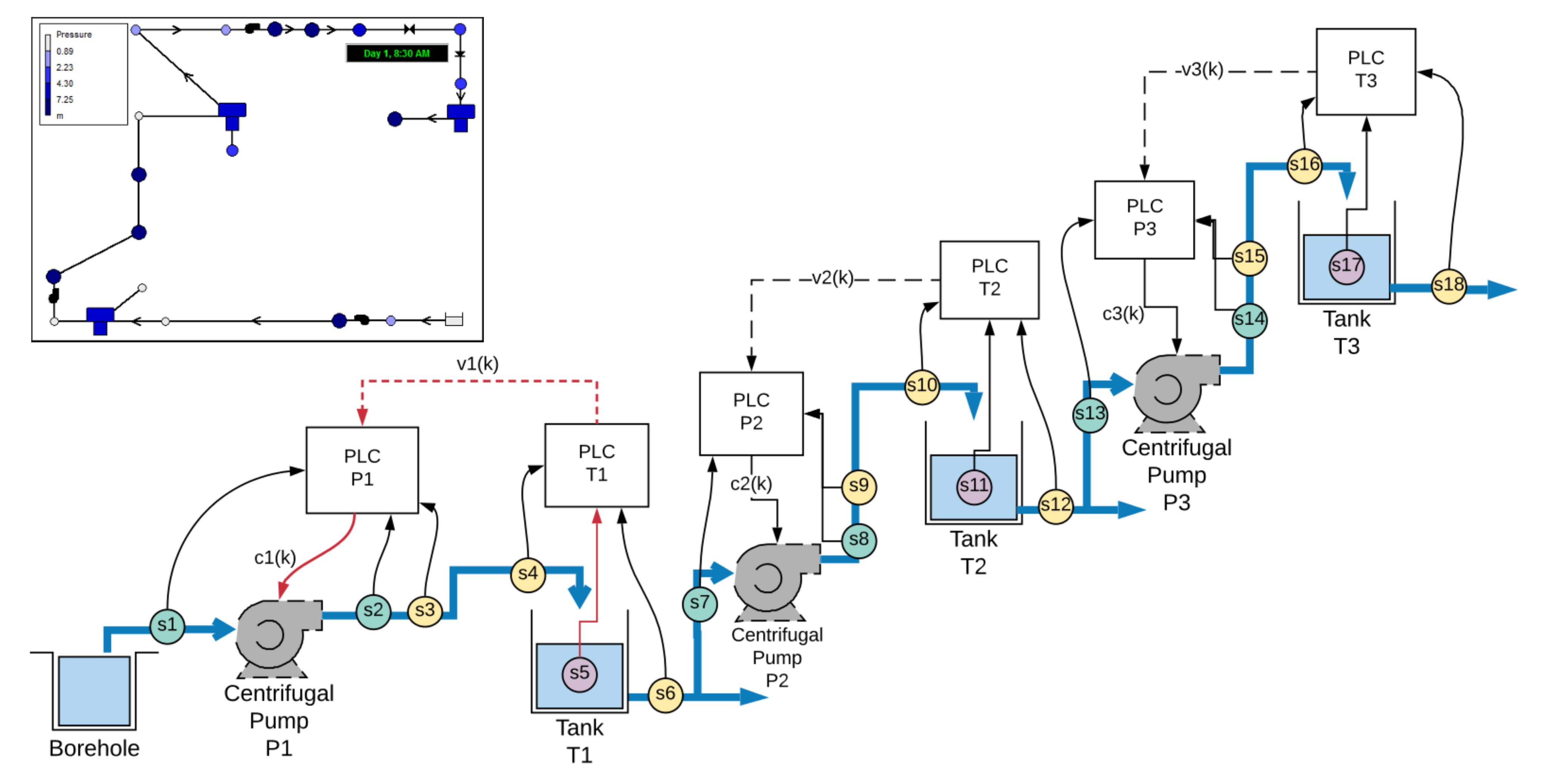}
	\caption{Three-tank water transport network architecture}
    \vspace{-0.5cm}
	\label{fig:3tank_wdn}
\end{figure}

The aforementioned basic component may appear repeatedly in larger networks. Figure \ref{fig:3tank_wdn} shows a subsystem of the network under analysis, which consists of three tanks in different elevations as well as three pumps that supply water to a different tank. Note that the subsystem from one pump to the next one is actually described by the structure of the basic subsystem presented in Figure \ref{fig:case-study-1}. Therefore, this example emphasizes that such basic structures are interesting on their own, as they can be combined to yield more complex subsystems in real WTNs. 
For the sake of brevity and simplicity, we focus here and thereafter on the study of basic structural subsystems as shown in \hbox{Figure \ref{fig:case-study-1}}.

\subsection{Data collection and initial scenario}

Various security measures are applied by water supply authorities in order to protect the components of their systems against malicious actors. 
Based on expert input provided by the operators that manage water transport networks with structures similar to the ones shown in Figures \ref{fig:case-study-1} and \ref{fig:3tank_wdn}, we have collected a list of measures that are used by water authorities to protect their network assets (i.e., sensors, flow meters, etc.). 
Using these measures, we have generated a security score for each network component of our case study as shown in Table~\ref{tab:secscores}. A more comprehensive explanation of these scores can be found in Annex \ref{annex:case-study}. Each score essentially measures the effort required by an attacker (i.e. attacker's cost) in order to compromise a component in the given configuration.

{\renewcommand{\arraystretch}{1.15}
    \begin{table}[!t]
        \martinTableFontsize
        \centering
        \begin{tabular}{|p{1.5cm}|p{0.5cm}|p{1.25cm}|p{3.9cm}|}
            \hline
            {\bf Components} & {\bf Cost} $\cost(n)$ & {\bf Measures} & { \bf Components state}\\
            \hline
            \centering $s6$		  & \centering 2 & $\{C, F\}$ & Stored in unlocked containers ($C$), placed in an open space in an unsecured fenced area ($F$). \\
            \hline
            \centering $s4$ & \centering 3 & $\{C, LC\}$ & Stored and locked  ($LC$) in containers  ($C$) in a public area. \\
            \hline
            \centering $s3$		  & \centering 5 & $\{C, F, AS\}$ & Stored in unlocked containers  ($C$), placed in an open space, with a fenced area ($F$) secured with alarm systems  ($AS$). \\
            \hline			
            \centering $a1, a3, a9$, $s5, s1$		& \centering 6 & $\{B, LB, F\}$ & Stored freely in a building  ($B$), which is locked  ($LB$), and placed in an unsecured fenced area ($F$). \\
            \hline			
            \centering $a2, a7, a8$, $a10, s2, c1$	& \centering 9 & $\{B, LB, F,$  $AS\}$ & Stored freely in a building ($B$), locked ($LB$), and placed in a secured fenced area ($F$) with alarm systems $(AS)$. \\
            \hline			
        \end{tabular}
        \vspace{0.1cm}
        \caption{Score table for physical security}
        \label{tab:secscores}
        \vspace{-0.8cm}
    \end{table}
}

The system components have been grouped in different categories according to their cost of compromise.  
Note that we do not provide a general methodology to derive a global scoring system; our scoring relies on the input we have collected and refers to the physical security of the components within the proposed scenario. 
Ideally, these scores should be combined with each component's cyber weaknesses by using, for example, CVSS scores \cite{CVSS}, in order to produce a more accurate scoring mechanism for cyber agents within ICS environments. 
To this day, however, there are no standard CVSS-like scoring systems specifically oriented to ICS networks. 

\begin{figure}[!h]
    \centering
    \includegraphics[scale=0.36]{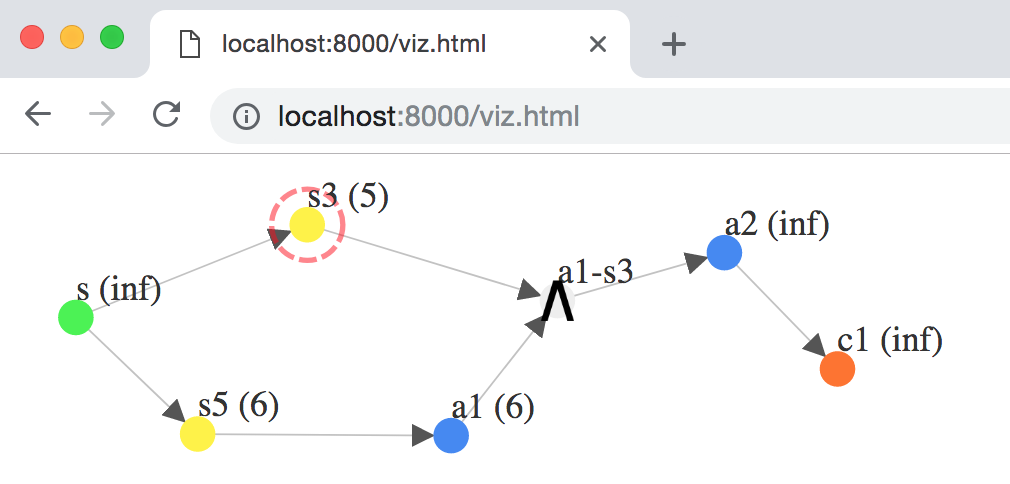} 
    \caption{Basic WTN without redundancy}
    \label{fig:wdn-simple}
\end{figure}

Given Table \ref{tab:secscores}, we first consider the basic scenario for our use case where redundant sensors are not used. Figure~\ref{fig:wdn-simple} illustrates the configuration and the solution of the graph derived from our use case for controlling actuator $c1$. 

In this scenario, only two agents exist, $a1$ and $a2$, and they function as previously explained. 
As $a2$ is the last agent that controls $c1$, any compromise of $a2$ or $c1$ would clearly disable the control of the system. In order to make the example more interesting, we consider a hypothetical situation where these components ($a2$ and $c1$) have been heavily protected and cannot be compromised (thus we assign an infinite cost to them), and we analyse the security level of the rest of the system. Since there are no other agents utilizing the measurements of sensors $s4$ and $s2$, these are not added in the graph as they do not contribute to the control of $c1$.

We have run our tool \tool over this configuration and the results indicate that the easiest component to compromise 
(so as to disrupt the operation of the system) is the flow sensor at the pump ($s3$) with a cost (effort) of 5, as shown in Figure \ref{fig:wdn-simple}.   
This is an interesting outcome since typical threats on these systems usually target tank level sensors ($s5$ in our case) in order to perform, for example, overflow and flooding attacks. 
However, our tool showed that, considering the specific settings collected from the plant, the physical security applied to sensor $s5$ exposes $s3$ as a more attractive target.

\subsection{Countermeasure recommendations and extended scenario}
\label{sec:counter}
By using the proposed security metric, new methodologies can be developed to enable analytical redundancies of different cyber-physical elements as an effort to reduce the risk of disrupting the system's monitoring and control operations. Redundancy is the act of introducing new hardware or software components to allow the generation of 
the same information through multiple sources. Typically, this can be achieved by adding new sensors 
(physical redundancy), or by using algorithms such as state estimates which compute a certain 
measurement using peripheral sensors (analytical redundancy).

As mentioned before, a critical component in our case study is the tank level sensor ($s5$). By compromising $s5$, an attacker can prevent the system from monitoring the level of the water in the tank. In such a case, the state of the system may be twofold: (i) the pump remains on, overflowing the tank, or (ii) the pump stays off and the tank empties at some point in time. With a similar reasoning, the flow sensor after the pump ($s3$) is another critical component since its failure will make PLC-P1 unaware of the pump status.

By adding the flow sensor $s6$ to measure the water outflow from the tank, and the pressure sensor $s1$ to measure the pressure of the water right before the pump, these sensors may allow the mitigation of the aforementioned issues. In particular, the tank's water level can be analytically estimated by combining the 
inflow measure at $s4$ and the outflow measure at $s6$, thus compensating for the water level measurement $s5$. 
On the other hand, monitoring the pressure at s1, together with s2, can be used to calculate the flow analytically, 
and therefore compensating for s3. 

\begin{figure}[h]
    \centering
    \includegraphics[scale=0.39]{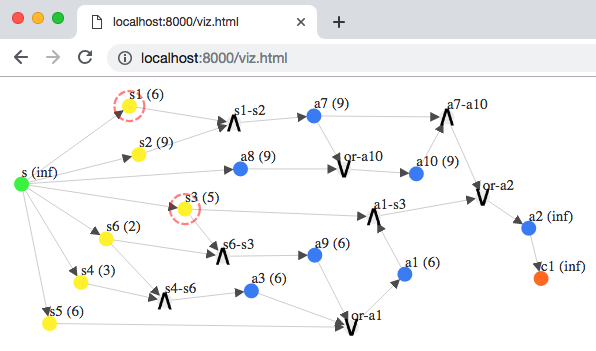} 
    \caption{Expanded WTN configuration with redundancy}
    \label{fig:wdn-expanded}
\end{figure}

These combinations generate new logical agents. In particular, $a3$ is the logical agent that is generated to accept 
the measurements from $s4$ and $s6$, and outputs the estimate of the water level at the tank. Moreover, agent $a7$ is added to accept the measurements from $s1$ and $s2$, and produces an estimate of the flow after the pump. Going a step further, the flow at $s3$ should be theoretically the same as the flow at $s4$. 
Therefore, the combination of $s3$ with $s6$ may yield agent $a9$ which, given these two measures, is able to compute an estimate of the level of the water in the tank. Finally, agents $a8$ and $a10$ may be derived from empirical results (e.g, what is the level in the tank based on the time that the pump is on) to estimate particular measurements. Using these redundancies we can derive the graph shown in Figure \ref{fig:wdn-expanded}.

After running \tool with the expanded configuration, we noticed that our metric still identifies the flow sensor at the pump ($s3$), along with the pressure sensor before the pump ($s1$), as the most critical components in the system. 
Interestingly, the water level sensor ($s5$) still does not constitute an attractive attack target. The added redundancy increased the cost of compromising the level measurement, and hence, a better strategy is to compromise the flow and the pressure sensors $\{s1, s3\}$ with cost 11. Such an attack will disable the control of the pump directly, despite of $a1$ being reporting the correct measurements of the water level at the tank to $a2$. Similarly, any other combination in the graph yields a cost higher than that of compromising the two aforementioned sensors. 
It should also be noted that the compromise cost of the expanded configuration (11) is much higher than the cost of compromising the basic configuration (5) in Figure~\ref{fig:wdn-simple}.

As a final remark, it is important to note that the initial scenario used in this section is reasonably small so as to properly explain its components and handle its complexity. Due to its size, its solution is also visible to the naked eye. 
However, when more components are added or its size increases, e.g. Fig. \ref{fig:wdn-expanded}, its solution is not that evident anymore. In that sense, our approach is intended to automate this process and assist administrators with actionable information that can be used to increase the security of their facilities.

\subsection{Further metric uses and potential applications}
In this section, we present other possible applications of our security metric over AND/OR graph models of ICS systems.

\begin{figure}[!t]
    \centering
    \includegraphics[scale=0.28]{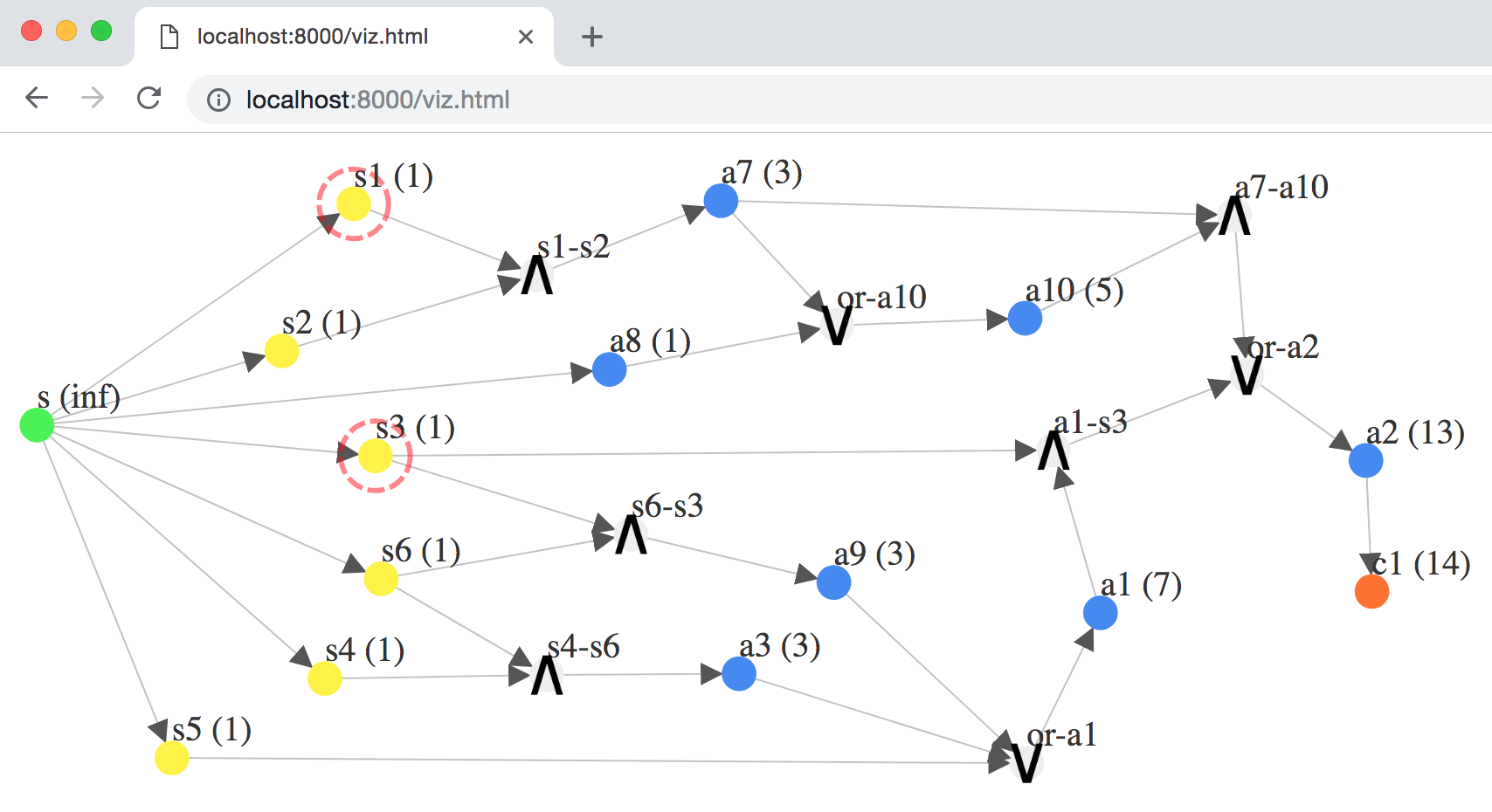}
    \caption{Security assignment for perimeter analysis}
    \label{fig:wdn-view1}
\end{figure}

\textbf{1) Perimeter analysis.}
The first application involves the security analysis of systems where the edge (or perimeter) nodes are easier to be compromised, and the effort becomes higher as we move to inner nodes of the graph.  
We can formulate such a system by assigning costs to the individual nodes based on their ``depth'' in a given graph $G$. Let $s$ be the source node in $G$, and $depth(n)$ a function to denote the depth of node $n$ in $G$, with $depth(s)=0$. Thus, we can define $depth(n)$ to be the minimum path from $s$ to $n$ in $G$, i.e. how many elements need to be visited (or compromised) before reaching $n$. The attack cost of $n$ can then be defined as $\cost(n) = depth(n)$. Note that we are changing the semantics of the cost function $\cost(n)$.  
Figure \ref{fig:wdn-view1} presents how this assignment can be applied to the components of our case study.

We observe that this cost assignment, combined to our security metric based on AND/OR graphs, provides a useful tool for the identification of the minimal set of critical edge components ($\{s1,s3\}$ in the above example) that need to be further protected to increase the security level of the system. 
The results for this specific configuration show that, as opposed to what one might initially think, not all of the nodes in the network perimeter are the most critical ones, but rather a subset of them. 
This showcases the fact that intricate dependencies among OT components (e.g. complex logical compositions) and specific control process flows establish a different perspective in how critical components must be understood. 
In that context, our security metric based on AND/OR graphs provides good support to properly identify and prioritise critical CPS components so as to improve the overall situational awareness.

\textbf{2) Prioritisation and threshold-based security hardening.}
The proposed security metric can also be used to gradually improve the security level of the network by prioritising CPS components that require utmost attention and applying remediation actions. 

In particular, the hardening process consists of a simple iteration where each stage determines the set of critical nodes in order to perform corrective tasks. Considering the example in Figure \ref{fig:wdn-view1}, let us remediate nodes $s1$ and $s3$ by setting an infinite cost to them. Which are now the new critical nodes? 
Computing the security metric with the updated settings, we obtain $\{s5,s6,s2\}$ as the new critical set with a total cost of 3. If we repeat the process after fixing $\{s5,s6,s2\}$, the critical set becomes $\{a1,a7\}$ with a total cost of 10, then $\{a2\}$ with cost 13, and finally $\{c1\}$ with cost 14. The same approach can be used to implement a threshold-based security hardening methodology where the process is repeated until the security level reaches the desired threshold.

\textbf{3) Budget-based security hardening.}
The previous threshold-based methodology assumes an infinite budget to remediate nodes. 
Sometimes, however, practical remediation costs are involved, in which case it is of interest to provide a balance between the efforts required by an attacker to compromise critical nodes and the cost required to fix them. 
Considering the proposed security metric, a simple way to achieve such balance is to add the remediation cost to the already defined cost function $\cost(n)$. This trade-off means that the cost of a node increases according to the complexity of compromise and how expensive its remediation is. In other words, an `easy' node to compromise will have a much larger value if it is too expensive to fix, and conversely, a complex node with almost no remediation cost will keep a similar score. 

More formally, we represent the budget required to remediate/fix a node $n$ as $\budget(n)$. 
We now redefine our cost function $\cost(n)$ as $\costprime(n)$ in order to reflect not only the cost for the attacker to compromise node $n$ but also the required budget to remediate it as follows:  
\begin{equation}
    \costprime_{\alpha,\beta}(n) = \alpha \times \cost(n) + \beta \times \budget(n)
\end{equation}
where $\alpha, \beta \in [0..1]$ are real coefficients such that $\alpha + \beta = 1$. 
The objective of these coefficients is to emphasis the importance of each factor within the new cost function $\costprime$. 
When $\alpha=0$ ($\costprime_{0,1}(n)$), the new cost function is completely focused on the budget. 
Conversely, when $\beta=0$ ($\costprime_{1,0}(n)$) , the new function gives us the original cost function focused on the attacker's cost. 
In that context, we can provide a balance between the two by parameterising the function with intermediate values $\alpha=0.5$ and $\beta=0.5$ ($\costprime_{0.5,0.5}$).  
A solution of our security metric with this new cost function will yield a set of critical nodes that can disrupt the operation of the target node while focusing on the remediation budget at the same time. 

In general terms, we have observed that the proposed security metric can be versatile enough to be used on different security applications. However, there are also other problems we plan to explore. For example, given a fixed budget $B$, what is the set of nodes that maximises the security level of the system within budget $B$? 
In addition, can our metric be used to support root-cause analysis in the context of incident response and digital forensics \cite{DFRWS2001, Stephenson2003}? These potential applications have already been scheduled as future work.

\section{Conclusions and future work}
\label{sec:conclusions-fw}

In this paper, we have presented a novel security metric aimed at identifying critical cyber-physical components and measuring the overall security of ICS environments. 
Central to our approach is the use of AND/OR graphs-based models, which allow more realistic representations of the complex interdependencies that are normally involved in real industrial networks. 
Based on these models, our security metric solves the problem of determining the set of ICS nodes that must be compromised in order to disrupt the operation of the system, with minimal cost for the attacker. 
Our strategy involves an efficient transformation of AND/OR graphs into weighted logical formulas that are then used to build and solve a Weighted Partial MAX-SAT problem. 
Our analytical experiments indicate that our computation strategy can properly scale to graphs with thousands of nodes in seconds. 
In addition, we have used our implementation prototype \tool to evaluate the security posture of a realistic water transport network, which has been presented in the form of a case study. 
During this study, we have identified interesting security properties about the plant as well as further security uses of our metric. 

As future work, we plan to extend our analysis on scalability and time variations by studying structural properties of AND/OR graphs over ICS networks, their computational time boundaries, and advanced graph-based techniques \cite{Barrere:CNSM2017}. 
To do so, we aim at further investigating with industrial partners what classes of graphs are typical and representative of different ICS networks. 
We also plan to extend our approach to consider multi-target attacks, defence budget constraints, and multiple overlapping on-site protection mechanisms. 
Automating the generation of input AND/OR graphs for ICS is also a challenging activity, which we plan to further investigate over real-world settings. 
In particular, we aim at a hybrid approach involving three main aspects: using semantic inference techniques to produce analytical redundancies~\cite{Millis2019}, IT-like network mapping and discovery mechanisms at the cyber level, and semi-automated methods to consolidate expert knowledge from operators. 
At the optimisation side, our computation strategy already considers a tie-break algorithm that selects the solution with minimum amount of nodes when two or more solutions with equal cost are found. 
However, in complex dense cases, deciding among minimal solutions with the same cost and the same amount of nodes requires further analysis. 
We also aim at studying the criticality of ICS components when nodes may be partially compromised or involve faulty signals, i.e., nodes that might partially operate under the presence of an attack (as opposed to be completely disconnected from the graph) \cite{DesmedtUsingFF}. 
Finally, we have shown that adding redundancy can increase the resiliency of an ICS environment. However, adding more components might also translate into an extended attack surface \cite{Li2018}. This aspect generates an interesting research problem whose solution might lead to a Pareto frontier regarding security levels and countermeasures.

\vspace{0.3cm}
\bibliographystyle{IEEEtran}
\bibliography{IEEEabrv,main-arXiv.bib}

\begin{appendices}
    \vspace{0.2cm}
\section{Tseitin transformation example}
\label{sec:annex-tseitin-example}

The Tseitin transformation \cite{Tseitin70} is an important technical part of our approach to produce equisatisfiable CNF formulas. From a practical perspective, we have experimentally observed that the naive CNF conversion method barely scales to a few hundred nodes before running out of memory. Instead, the Tseitin transformation  enables our approach to scale to graphs with thousand of nodes in a matter of seconds. In this section, we exemplify how the Tseitin transformation works and the kind of formulas we obtain from it. 
Let us consider the following logical formula: 
\[
\phi = (p \lor q) \land r
\]

The subformulas involved in $\phi$ (non-atomic) are: 

\begin{enumerate}[i.]
    \item $p \lor q $
     \item $(p \lor q) \land r$
\end{enumerate}

We now introduce a new variable for each subformula as follows: 

\begin{enumerate}[i.]
    \item $x_1 \leftrightarrow p \lor q $
    \item $x_2 \leftrightarrow x_1 \land r$ (note that we are using $x_1$ instead of $p \lor q$)
\end{enumerate}

Putting all substitutions together (including $x_2$ as the substitution of $\phi$), we obtain the following transformed formula: 
\[
\tau(\phi) = x_2 \land (x_2 \leftrightarrow x_1 \land r) \land (x_1 \leftrightarrow p \lor q) 
\]

Now each conjunct in $\tau(\phi)$ can be individually converted to its conjunctive normal form (CNF). 
For $x_1 \leftrightarrow p \lor q$, we have that: 
\begin{align*} 
x_1 \leftrightarrow p \lor q&\equiv  (x_1 \rightarrow (p \lor q)) \land ((p \lor q) \rightarrow x_1)\\ 
&\equiv  (\neg x_1 \lor p \lor q) \land (\neg (p \lor q) \lor x_1)\\ 
&\equiv  (\neg x_1 \lor p \lor q) \land ((\neg p \land \neg q) \lor x_1)\\
&\equiv  (\neg x_1 \lor p \lor q) \land (\neg p \lor x_1) \land (\neg q \lor x_1)
\end{align*}
It can be observed that after applying a few logical equivalence rules, the obtained formula is in CNF. 
For $x_2 \leftrightarrow x_1 \land r$, we have that: 
\begin{align*} 
x_2 \leftrightarrow x_1 \land r &\equiv  (x_2 \rightarrow (x_1 \land r)) \land ((x_1 \land r) \rightarrow x_2)\\ 
&\equiv  (\neg x_2 \lor (x_1 \land r)) \land (\neg (x_1 \land r) \lor x_2)\\ 
&\equiv  (\neg x_2 \lor x_1) \land (\neg x_2 \lor r) \land (\neg (x_1 \land r) \lor x_2)\\ 
&\equiv  (\neg x_2 \lor x_1) \land (\neg x_2 \lor r) \land (\neg x_1 \lor \neg r \lor x_2)
\end{align*}
Finally, substituting each clause in $\tau(\phi)$  by its corresponding CNF conversion as shown before, we obtain a new CNF formula with additional variables that is not equivalent to the original one, but it is equisatisfiable. This means that for any truth assignment, $\phi$ and $\tau(\phi)$ will always be either both $true$ or both $false$. 
The expanded new CNF formula is as follows: 
\begin{align*} 
\tau(\phi) = & x_2 \land  (\neg x_1 \lor p \lor q) \land (\neg p \lor x_1) \land (\neg q \lor x_1) \land \\
& (\neg x_2 \lor x_1) \land (\neg x_2 \lor r) \land (\neg x_1 \lor \neg r \lor x_2)
\end{align*}

    \vspace{0.2cm}
\section{Handling cycles in AND/OR graphs}
\label{sec:annex-cycles}

\begin{figure}[h]
    \centering
    \includegraphics[scale=0.27]{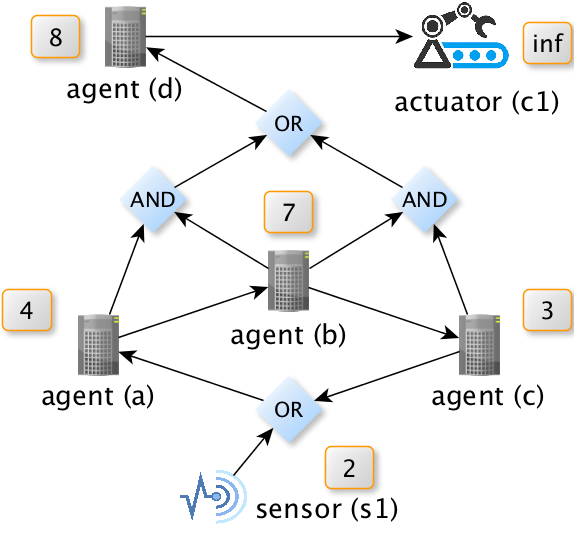}
    \caption{Cycle example (between $a$, $b$, and $c$)}
    \label{fig:cycle-example}
\end{figure}

Let us consider the graph illustrated in Figure \ref{fig:cycle-example}. 
\vspace{-0.25cm}

\vspace{-0.2cm} 
As explained in Section \ref{sec:computation-strategy}, the logical transformation $\form(c1)$ starts at a target node $c1$ and traverses the graph backwards until all the components in the graph have been covered.
It is easy to see that at some point, the partial formula in this example will have the following aspect (where nodes $a$, $b$ and $c$ have not yet been expanded): 
\vspace{-0.05cm}
\[
\form(c1) = c1 \land d \land ((a \land b) \lor (b \land c))
\]

When node $a$ is expanded with the nodes it depends on, we can see that $a$ depends on $(s1 \lor c)$. While $s1$ is a terminal node and does not depend on any node, node $c$ still depends on $b$, and $b$ eventually depends on $a$. 
Because node $a$ was already visited, the transformation process stops the exploration at this point (does not go further from $a$ again), with the following partial formula: 
\vspace{-0.2cm}
\[
\form(c1) = c1 \land d \land (((a \land (s1 \lor (c \land b \land a))) \land b) \lor (b \land c))
\]

\begin{figure}[h]
    \centering
    \includegraphics[scale=0.307]{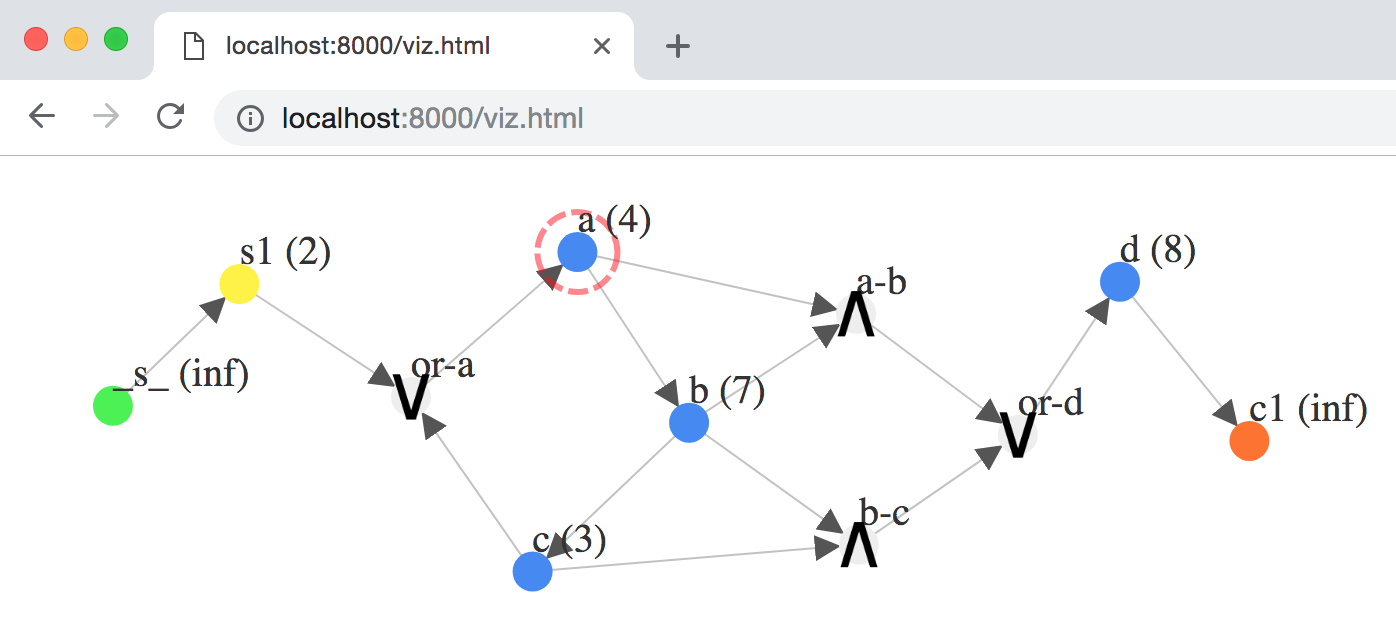}
    \caption{Cycle example (between $a$, $b$, and $c$) - \tool viewer}
    \label{fig:cycle-example-view}
\end{figure}

Clearly, the last $a$ can be removed since it already appears early in the formula as a predecessor connected with the AND operator. Hypothetically, if we follow the loop indefinitely, we would see the same pattern again and again. From a satisfiability perspective, it would yield the same result since the same variables are conjunctively joined in the sentence.  A similar situation occurs when nodes $b$ and $c$ are expanded. At the end, the final transformation looks as follows: 
\begin{align*} 
\form(c1) = &c1 \land d \land (((a \land (s1 \lor (c \land b))) \land (b \land a \land (s1 \lor c)) ) \\
& \lor ( (b \land a \land (s1 \lor c)) \land (c \land b \land a \land s1) ))
\end{align*}
The metric resolution for this graph is shown in \hbox{Figure \ref{fig:cycle-example-view}}.

\vspace{-0.2cm}

It is important to note that a typical approach to deal with cycles, widely used in many graph-related works, is to analyse each cycle as a whole. 
That is, treat loopy formations as clusters that can be collapsed and analysed as one super node where its cost is equal to the minimum cost among its member nodes. From a graph-theoretical perspective, these clusters are strongly connected components (SCC) and can be efficiently identified in linear time using, for example, Tarjan's algorithm~\cite{Tarjan1972}. However, such an approach does not properly work with AND/OR graphs. The previous scenario is a counterexample. The minimum node cost within the cycle is 3 (node $c$), however, the solution to that problem is node $a$ with cost 4. This is because the dependencies outside the loop may affect the overall optimal solution.

    \vspace{0.2cm}
\section{\tool execution over example scenario}
\label{sec:tool-execution}

\begin{figure}[t]
    \vspace{-0.35cm}
\begin{lstlisting}[
label={lst:json_input}, 
mathescape=true, language=json, caption=Graph specification - JSON input, 
basicstyle=\martinListingCodeFontsize\sffamily, 
keywordstyle=\bfseries\color{green!40!black}, 
commentstyle=\itshape\color{purple!40!black},
identifierstyle=\color{blue},
stringstyle=\color{orange}, 
frame=single, 
%numbers=left,
numberstyle=\tiny\color{gray},
stepnumber=1, 
belowcaptionskip=5em,
belowskip=3em, 
captionpos=b
]
{
    "graph": {
        "target":"c1", 
        "nodes": [
            { "id": "c1", "type": "actuator", "value": "inf" },
            { "id": "d", "type": "agent", "value": "10" }, 
            { "id": "or-d", "type": "or", "value": "none" }, 
            { "id": "c", "type": "sensor", "value": "2" },
            { "id": "b", "type": "agent", "value": "5" },
            { "id": "a", "type": "sensor", "value": "2" },
            { "id": "a-b", "type": "and", "value": "none" },
            { "id": "b-c", "type": "and", "value": "none" }
        ],
        "edges": [
            { "source": "d", "target": "c1" }, 
            { "source": "or-d", "target": "d" },
            { "source": "a-b", "target": "or-d" }, 
            { "source": "b-c", "target": "or-d" }, 
            { "source": "a", "target": "a-b" },
            { "source": "b", "target": "a-b" },
            { "source": "b", "target": "b-c" },
            { "source": "c", "target": "b-c" }
        ]
    }
} 
\end{lstlisting}
\vspace{-0.65cm}
\end{figure}

Listing \ref{lst:json_input} depicts the JSON specification file for our example scenario, initially described in Section \ref{sec:model}, Figure \ref{fig:simple-example1}. The graph encoding essentially involves a list of nodes, the edges among them, and the target node under analysis. 

Once the input graph has been pre-processed as explained in Section \ref{sec:impl} (step 2), the graph is translated into a logical representation (step 3) that is later transformed into an instance of a Weighted Partial MAX-SAT problem (step 4). Listing~\ref{lst:example1-exec1} shows a partial output of \tool when it is executed with the previous JSON specification file (Listing \ref{lst:json_input}), which indicates node $c1$ as the target node. 

\begin{figure}[!h]
    \vspace{-0.25cm}
\begin{lstlisting}[
label={lst:example1-exec1}, 
mathescape=true, language=bash, caption=Security metric resolution, 
basicstyle=\martinListingCodeFontsize\sffamily, 
keywordstyle=\bfseries\color{green!40!black}, 
commentstyle=\itshape\color{purple!40!black},
identifierstyle=\color{blue},
stringstyle=\color{orange}, 
frame=single, 
%numbers=left,
numberstyle=\tiny\color{gray},
stepnumber=1, 
belowcaptionskip=5em,
belowskip=3em, 
captionpos=b
]
 Logical formula: 
 c1 & ( d & ( ( ( a & s ) & ( b & s ) ) | ( ( b & s ) & ( c & s ) ) ) )
 
 Objective: 
 $\sim$( c1 & ( d & ( ( ( a & s ) & ( b & s ) ) | ( ( b & s ) & ( c & s ) ) ) ) )
 
 Tseitin CNF sentence (DIMACS): 
 - Number of variables: 15
 - Number of clauses: 27
 
==================================
### BEST solution found: 
=== Security Metric ===
CUT cost: 4.0
CUT solution: (c,2); (a,2); 
\end{lstlisting}
\vspace{-0.65cm}
\end{figure}

In mathematical terms, the first logical formula corresponds to $\form(c1)$, which indicates the means to fulfil the target node $c1$.  
The second formula is the objective of the attacker, which mathematically corresponds to $\neg \form(c1)$. The third part (not fully displayed) shows the obtained CNF formula after applying the Tseitin transformation. Our example has 15 logical variables and 27 clauses. 
We use the DIMACS format to display the CNF formula \cite{DIMACS}. Finally, the Weighted Partial MAX-SAT problem is built using the appropriate clauses and costs, as described in Section \ref{sec:sat-formulation}. 

In this scenario, the solution obtained with the MAX-SAT solver (step 4) indicates that the best strategy for the attacker is to compromise nodes $a$ and $c$ with a minimal total cost of 4.0. 
\tool outputs another JSON file that includes the original graph and also specifies the minimum vertex cut as shown in Listing \ref{lst:json_solution} (step 5). 

\begin{figure}[!h]
    \vspace{-0.15cm}
\begin{lstlisting}[
label={lst:json_solution}, 
mathescape=true, language=json, caption=JSON output with solution - Critical nodes and total cost, 
basicstyle=\martinListingCodeFontsize\sffamily, 
keywordstyle=\bfseries\color{green!40!black}, 
commentstyle=\itshape\color{purple!40!black},
identifierstyle=\color{blue},
stringstyle=\color{orange}, 
frame=single, 
%numbers=left,
numberstyle=\tiny\color{gray},
stepnumber=1, 
belowcaptionskip=5em,
belowskip=3em, 
captionpos=b
]
{
    "graph": {
        ..., 
        "nodes": [ ... ], 
        "edges": [ ... ]
    },
    "cut" : {
        "nodes" : [ 
            { "id" : "a", "type" : "sensor", "value" : "2" }, 
            { "id" : "c", "type" : "sensor", "value" : "2" } 
        ],
        "cost" : 4.0
    }	
}
\end{lstlisting}
\vspace{-0.65cm}
\end{figure}

The output JSON depicted in Listing \ref{lst:json_solution} is then used to feed the visualisation component (step 6). 
Based on the information provided in the vertex cut section, the system displays the critical nodes as shown in Figure \ref{fig:example1-viewer}, Section \ref{sec:impl} (surrounded by dashed red circles), and allows the user to validate the solution by interactively removing them until the target is disabled.  

It is worth noting that when different costs are considered, the optimal solution might naturally change. If we now consider a different cost for node $b$, for example $\cost(b) = 3.2$, we obtain the results depicted in Listing \ref{lst:example1-exec2}, and consequently, a new JSON file including the new security assessment results. 

\begin{lstlisting}[
label={lst:example1-exec2},
caption=\text{Security metric resolution with $\cost(b)=3.2$}, 
mathescape=true, language=bash, 
basicstyle=\martinListingCodeFontsize\sffamily, 
keywordstyle=\bfseries\color{green!40!black}, 
commentstyle=\itshape\color{purple!40!black},
identifierstyle=\color{blue},
stringstyle=\color{orange}, 
frame=single, 
%numbers=left,
numberstyle=\tiny\color{gray},
stepnumber=1, 
belowcaptionskip=5em,
belowskip=3em, 
captionpos=b
]
 Logical formula: 
 c1 & ( d & ( ( ( a & s ) & ( b & s ) ) | ( ( b & s ) & ( c & s ) ) ) )

 Objective: 
 $\sim$( c1 & ( d & ( ( ( a & s ) & ( b & s ) ) | ( ( b & s ) & ( c & s ) ) ) ) )

 Tseitin CNF sentence (DIMACS): 
 - Number of variables: 15
 - Number of clauses: 27

==================================
### BEST solution found: 
=== Security Metric ===
CUT cost: 3.2
CUT solution: (b,3.2); 
\end{lstlisting}

\begin{figure}[h]
	\centering
	\includegraphics[scale=0.47]{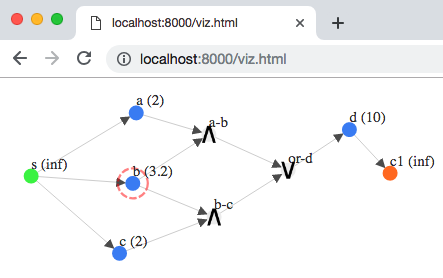}
	\caption{Graph viewer - metric resolution with $\cost(b)=3.2$}
	\label{fig:example1b-viewer}
\end{figure}

Figure~\ref{fig:example1b-viewer} shows the results with the \tool graph viewer. 
 
    \section{Case study - Score tables}
\label{annex:case-study}

\begin{table}[!h]
    \martinTableFontsize
	\centering
	\begin{tabular}{|p{1.2cm}|p{1.25cm}|c|p{4cm}|}
		\hline
		{\bf Num.} & {\bf Notation} & {\bf Weight} & {\bf Measure}\\
		\hline
		\hline
		1 & C & 1 & Stored In Container. \\
		\hline
		2 & LC & 2 & Locked Container \\
		\hline
		3 & F &  1 & Fenced Area \\
		\hline
		4 & B & 2 & Stored In Building \\
		\hline
		5 & LB & 3 & Locked Building \\
		\hline
		6 & AS & 3 & Monitored Area (Alarm System) \\
		\hline
		7 & MA & 4 & Multi-Authentication \\
		\hline
	\end{tabular}
	\vspace{0.1cm}
	\caption{Security Measures to protect assets}
	\label{tab:measures}
	\vspace{-0.5cm}
\end{table}

Various security measures are applied by water supply authorities in order to protect the components of their systems against malicious actors. 
Table \ref{tab:measures} presents a sample list of measures taken by the authorities. Let each measure $m$ have a weight $w_m$, which is based on the level of difficulty an attacker will have to deal with it. Also let $M_n$ be the set of measures taken for a component $n$. Thus, we compute the security score of a component $n$ as:
\[
\cost(n) = \sum_{m\in M_n}w_m
\]
\vspace{0.1cm}

In Table \ref{tab:scores}, we enumerate the possible ``physical'' configurations for the components within our example system, and provide a security score which measures the effort that an attacker should make (i.e. attacker's cost) in order to compromise a component in the given configuration. The system components have been grouped in different categories according to their cost of compromise. It is important to note that this scoring relies on the input we have collected and refers to the physical security of the components within the proposed scenario. 

\vspace{0.1cm}
{\renewcommand{\arraystretch}{1.15}
\begin{table}[!h]
	\centering
	{\scriptsize
		\begin{tabular}{|p{1.3cm}|p{0.5cm}|p{1.9cm}|p{3.6cm}|}
			\hline
			{\bf Components} & {\bf Cost} $\cost(n)$ & $M_n$ & { \bf Components state}\\
			\hline
			& 0 & $\{\}$ &Open to the air and access is not limited by any means. \\
			\hline
			& 1 & $\{C\}$ &Stored in an unlocked container ($C$) in a public area, with uncontrolled access. \\
			\hline
			\centering $s6$		  & 2 & $\{C, F\}$ & Stored in unlocked containers ($C$), placed in an open space in an unsecured fenced area ($F$). \\
			\hline
			\centering$s4$ & 3 & $\{C, LC\}$ & Stored and locked  ($LC$) in containers  ($C$) in a public area. \\             
			\hline
			\centering$s3$		  & 5 & $\{C, F, AS\}$ & Stored in unlocked containers  ($C$), placed in an open space, with a fenced area ($F$) secured with alarm systems  ($AS$). \\
			\hline
			\centering$a1, a3, a9$, $s5, s1$		& 6 & $\{B, LB, F\}$ & Stored freely or unlocked in a building  ($B$), which is locked  ($LB$), and placed in an unsecured fenced area ($F$). \\
			\hline
			\centering$a2, a7, a8$, $a10, s2, c1$	& 9 & $\{B, LB, F, AS\}$ & Stored freely in a building ($B$), locked ($LB$), and placed in a secured fenced area ($F$) with alarm systems $(AS)$. \\
			\hline
			& 9 & $\{C, LC, B, LB$, $F\}$ & Stored in containers ($C$) with a lock mechanism ($LC$), inside a building ($B$), which is locked ($LB$), and placed in an unsecured fenced area ($F$). \\
			\hline
			& 12 & $\{C, LC, B, LB$, $F, AS\}$ & Stored in containers ($C$) with a lock mechanism ($LC$), inside a building ($B$), which is locked ($LB$), and placed in a secured fenced area ($F$) with alarm systems ($AS$). \\
			\hline
			& 16 & $\{C, LC, B, LB$, $F, AS, MA\}$ & Stored in containers ($C$) with a lock mechanism ($LC$), inside a building ($B$), which is locked ($LB$) with a multi-authentication system ($MA$), and placed in a secured fenced area ($F$) with alarm systems ($AS$). \\
			\hline		
		\end{tabular}
	}
	\vspace{0.2cm}
	\caption{Score table for physical security}
	\label{tab:scores}
	\vspace{-0.36cm}
\end{table}
}

\end{appendices}

\end{document}